\documentclass[twocolumn]{aastex6}
\pdfoutput=1
\usepackage{fancyhdr}
\usepackage{color}
\usepackage{natbib}
\usepackage{amssymb,amsmath}
\usepackage{graphicx}
\usepackage{multirow}
\usepackage[normalem]{ulem}
\usepackage[utf8]{inputenc}
\usepackage{wrapfig}
\shorttitle{}
\shortauthors{Chilcote et al.}

\newcommand{\teff}{$T_\mathrm{eff}$}
\newcommand{\logg}{$\log g$~}
\newcommand{\rjup}{$\mathcal{R}_\mathrm{Jup}$}
\newcommand{\mjup}{$\mathcal{M}_\mathrm{Jup}$}

\begin{document}

\title{1 to 2.4 micron Near-IR spectrum of the Giant Planet $\beta$ Pictoris \MakeLowercase{b} \\ obtained with the Gemini Planet Imager}


\author{Jeffrey Chilcote\altaffilmark{1}, Laurent Pueyo\altaffilmark{2}, Robert J. De Rosa\altaffilmark{3}, Jeffrey Vargas\altaffilmark{3}, Bruce Macintosh\altaffilmark{4}, Vanessa P. Bailey\altaffilmark{4}, Travis Barman\altaffilmark{5}, Brian Bauman\altaffilmark{6}, Sebastian Bruzzone\altaffilmark{7}, Joanna Bulger\altaffilmark{8}, Adam S. Burrows\altaffilmark{9}, Andrew Cardwell\altaffilmark{10,11}, Christine H. Chen\altaffilmark{2}, Tara Cotten\altaffilmark{12}, Daren Dillon\altaffilmark{13}, Rene Doyon\altaffilmark{14}, Zachary H. Draper\altaffilmark{15,16}, Gaspard Duch{\^e}ne\altaffilmark{3,17}, Jennifer Dunn\altaffilmark{16}, Darren Erikson\altaffilmark{16}, Michael P. Fitzgerald\altaffilmark{18}, Katherine B. Follette\altaffilmark{4,35}, Donald Gavel\altaffilmark{13}, Stephen J. Goodsell\altaffilmark{19,20}, James R. Graham\altaffilmark{3}, Alexandra Z. Greenbaum\altaffilmark{21}, Markus Hartung\altaffilmark{10}, Pascale Hibon\altaffilmark{22}, Li-Wei Hung\altaffilmark{18}, Patrick Ingraham\altaffilmark{23}, Paul Kalas\altaffilmark{3,24}, Quinn Konopacky\altaffilmark{25}, James E. Larkin\altaffilmark{18}, J{\'e}r{\^o}me Maire\altaffilmark{25}, Franck Marchis\altaffilmark{24}, Mark S. Marley\altaffilmark{26}, Christian Marois\altaffilmark{15,16}, Stanimir Metchev\altaffilmark{7}, Maxwell A. Millar-Blanchaer\altaffilmark{27,36}, Katie M. Morzinski\altaffilmark{28}, Eric L. Nielsen\altaffilmark{4,24}, Andrew Norton\altaffilmark{13}, Rebecca Oppenheimer\altaffilmark{29}, David Palmer\altaffilmark{6}, Jennifer Patience\altaffilmark{30}, Marshall Perrin\altaffilmark{2}, Lisa Poyneer\altaffilmark{6}, Abhijith Rajan\altaffilmark{30}, Julien Rameau\altaffilmark{14}, Fredrik T. Rantakyr{\"o}\altaffilmark{10}, Naru Sadakuni\altaffilmark{31}, Leslie Saddlemyer\altaffilmark{16}, Dmitry Savransky\altaffilmark{32}, Adam C. Schneider\altaffilmark{30}, Andrew Serio\altaffilmark{10}, Anand Sivaramakrishnan\altaffilmark{2}, Inseok Song\altaffilmark{12}, Remi Soummer\altaffilmark{2}, Sandrine Thomas\altaffilmark{23}, J. Kent Wallace\altaffilmark{27}, Jason J. Wang\altaffilmark{3}, Kimberly Ward-Duong\altaffilmark{30}, Sloane Wiktorowicz\altaffilmark{33}, and Schuyler Wolff\altaffilmark{34}}

\altaffiltext{1}{Dunlap Institute for Astronomy \& Astrophysics, University of Toronto, Toronto, ON M5S 3H4, Canada}
\altaffiltext{2}{Space Telescope Science Institute, Baltimore, MD 21218, USA}
\altaffiltext{3}{Astronomy Department, University of California, Berkeley; Berkeley CA 94720, USA}
\altaffiltext{4}{Kavli Institute for Particle Astrophysics and Cosmology, Department of Physics, Stanford University, Stanford, CA, 94305, USA}
\altaffiltext{5}{Lunar and Planetary Laboratory, University of Arizona, Tucson AZ 85721, USA}
\altaffiltext{6}{Lawrence Livermore National Laboratory, 7000 East Ave., Livermore, CA 94550}
\altaffiltext{7}{Department of Physics and Astronomy, Centre for Planetary Science and Exploration, the University of Western Ontario, London, ON N6A 3K7, Canada}
\altaffiltext{8}{Subaru Telescope, NAOJ, 650 North A'ohoku Place, Hilo, HI 96720, USA}
\altaffiltext{9}{Department of Astrophysical Sciences, Princeton University, Princeton, NJ 08544, USA}
\altaffiltext{10}{Gemini Observatory, Casilla 603, La Serena, Chile}
\altaffiltext{11}{Large Binocular Telescope Observatory, 933 N Cherry Ave., Tucson AZ 85721, USA}
\altaffiltext{12}{Department of Physics and Astronomy, University of Georgia, Athens, GA 30602, USA}
\altaffiltext{13}{University of California Observatories/Lick Observatory, University of California, Santa Cruz;  Santa Cruz, CA 95064, USA}
\altaffiltext{14}{Institut de Recherche sur les Exoplan{\`e}tes, D{\'e}partment de Physique, Universit{\'e} de Montr{\'e}al, Montr{\'e}al QC H3C 3J7, Canada}
\altaffiltext{15}{University of Victoria, 3800 Finnerty Rd, Victoria, BC V8P 5C2, Canada}
\altaffiltext{16}{National Research Council of Canada Herzberg, 5071 West Saanich Rd, Victoria, BC V9E 2E7, Canada}
\altaffiltext{17}{Univ. Grenoble Alpes/CNRS, IPAG, F-38000 Grenoble, France}
\altaffiltext{18}{Department of Physics \& Astronomy, University of California, Los Angeles, CA 90095, USA}
\altaffiltext{19}{Gemini Observatory, 670 N. A'ohoku Place, Hilo, HI 96720, USA}
\altaffiltext{20}{Department of Physics, Durham University, Stockton Road, Durham DH1, UK}
\altaffiltext{21}{Department of Astronomy, University of Michigan, Ann Arbor, MI 48109, USA}
\altaffiltext{22}{European Southern Observatory, Alonso de Cordova 3107, Vitacura, Santiago, Chile}
\altaffiltext{23}{Large Synoptic Survey Telescope, 950N Cherry Av, Tucson, AZ 85719, USA}
\altaffiltext{24}{SETI Institute, Carl Sagan Center, 189 Bernardo Avenue,  Mountain View, CA 94043, USA}
\altaffiltext{25}{Center for Astrophysics and Space Science, University of California San Diego, La Jolla, CA 92093, USA}
\altaffiltext{26}{Space Science Division, NASA Ames Research Center, Mail Stop 245-3, Moffett Field CA 94035, USA }
\altaffiltext{27}{Jet Propulsion Laboratory, California Institute of Technology Pasadena CA 91125, USA}
\altaffiltext{28}{Steward Observatory, University of Arizona, Tucson AZ 85721, USA}
\altaffiltext{29}{American Museum of Natural History, Department of Astrophysics, New York, NY 10024, USA}
\altaffiltext{30}{School of Earth and Space Exploration, Arizona State University, PO Box 871404, Tempe, AZ 85287, USA}
\altaffiltext{31}{Stratospheric Observatory for Infrared Astronomy, Universities Space Research Association, NASA/Armstrong Flight Research Center, 2825 East Avenue P, Palmdale, CA 93550, USA}
\altaffiltext{32}{Sibley School of Mechanical and Aerospace Engineering, Cornell University, Ithaca, NY 14853, USA}
\altaffiltext{33}{The Aerospace Corporation, 2310 E. El Segundo Blvd., El Segundo, CA 90245}
\altaffiltext{34}{Department of Physics and Astronomy, Johns Hopkins University, Baltimore, MD 21218, USA}
\altaffiltext{35}{NASA Sagan Fellow}
\altaffiltext{36}{NASA Hubble Fellow}


\begin{abstract}

Using the Gemini Planet Imager (GPI) located at Gemini South, we measured the near-infrared (1.0--2.4\,\micron) spectrum of the planetary companion to the nearby, young star $\beta$~Pictoris. 
We compare the spectrum obtained with currently published model grids and with known substellar objects and present the best matching models as well as the best matching observed objects. Comparing the empirical measurement of the bolometric luminosity to evolutionary models, we find a mass of $12.9\pm0.2$\,\mjup, an effective temperature of $1724\pm15$\,K, a radius of $1.46\pm0.01$\,\rjup, and a surface gravity of $\log g = 4.18\pm0.01$\,[dex] (cgs). The stated uncertainties are statistical errors only, and do not incorporate any uncertainty on the evolutionary models. Using atmospheric models, we find an effective temperature of $1700-1800$\,K and a surface gravity of $\log g = 3.5$--$4.0$\,[dex] depending upon model. These values agree well with other publications and with ``hot-start'' predictions from planetary evolution models. Further, we find that the spectrum of $\beta$~Pic~b best matches a low-surface gravity L2$\pm$1 brown dwarf. Finally comparing the spectrum to field brown dwarfs we find the the spectrum best matches 2MASS J04062677--381210 and 2MASS J03552337+1133437.

\end{abstract}


\keywords{(stars:beta Pictoris) planetary systems --- instrumentation: adaptive optics --- techniques: spectroscopic --- infrared: general}


\section{Introduction}

Since the discovery of 51~Pegasi~b in 1995 \citep{MayorQueloz1995}, the search for and discovery of extrasolar planets has broadly changed our understanding of planetary systems. Direct imaging allows for the discovery of planets on solar system-scale orbits, provides new insight into the formation and characteristics of extrasolar systems, and enable direct spectroscopic observations of their atmospheres. Despite decades of efforts to image young Jupiter-mass exoplanets still luminous as a result of their formation process, only a handful of extrasolar planets have ever been directly imaged. Examples of such planets include 2M1207b \citep{Chauvin2005}, \mbox{Fomalhaut b} \citep{Kalas2008}, the HR8799 system \citep{Marois2008,Marois2010}, $\beta$~Pic~b \citep{Lagrange2010}, \mbox{IRXS J1609 b} \citep{Lafreniere2010}, \mbox{HD 95086 b} \citep{Rameau2013}, 51 Eri b \citep{Macintosh2015}, and HD 131399 Ab \citep{2016Sci...353..673W}.

$\beta$~Pictoris ($\beta$~Pic, HD 39060) is a $24\pm3$\,Myr \citep{Bell2015}, A6V star located at a distance of $19.44\pm0.05$\,pc \citep{Gray2006, vanLeeuwen2007}. $\beta$~Pic represents the earliest example of high contrast imaging to directly detect a circumstellar disk \citep{SmithTerrile1984}. The disk is seen edge-on and shows an asymmetric structure that has been attributed to planetary perturbations  \citep{Burrows1995,Kalas1995,Golimowski2006,Mouillet1997,Heap2000,Augereau2001}. The planet $\beta$~Pic~b was first detected by VLT/NaCo \citep{Lagrange2010}. Since then, its orbit has been constrained via careful astrometric monitoring \citep{chauvin2012,Nielsen2014,MillarBlanchaer2015,wang2016}. The atmospheric properties of the planet have been estimated from photometric and spectroscopic measurements using a number of adaptive optics (AO) fed instruments such as Gemini/NICI \citep{Boccaletti2013}, Magellan AO \citep{Males2014,Morzinski2015}, Gemini/GPI \citep{Chilcote2015}, and VLT/SPHERE \citep{Baudino2015}. 

 Because of the presence of a dynamically perturbed debris disk \citep{Mouillet1997,2012A&A...542A..40L,MillarBlanchaer2015}, along with well documented constraints on its age \citep{Bell2015}, the $\beta$~Pic planetary system is an ideal laboratory to understand the formation and evolution of sub-stellar objects near the planet/brown dwarf limit. The luminosity and colors of $\beta$~Pic~b are indeed similar to early-type brown dwarfs \citep{Males2014,Morzinski2015,Bonnefoy2014,Currie2013}. However, constraints from radial velocity observations place its dynamical mass well below the value expected for an isolated field object of the same luminosity \citep{Lagrange2012}. The recently identified population of young isolated brown dwarfs with low-surface gravity \citep{Kirkpatrick:2008ec,Allers2013,2009AJ....137.3345C,2012A&A...548A..26D,Faherty:2013bc, Gagne:2015dc, Schneider:2016iq} is a more appropriate sample to compare to $\beta$~Pic~b. Recent work has provided a preliminary look at the near-infrared low resolution spectrum of $\beta$~Pic~b \citep{Bonnefoy2014, Chilcote2015, Baudino2015} and has highlighted the similarities between $\beta$~Pic~b and low-gravity brown dwarfs in young associations and moving groups. Such results naturally lead to questions regarding the formation mechanisms underlying these two type of objects that have apparently very different dynamical origins (isolated vs. orbiting another star) and yet look similar from a spectro-photometric standpoint \citep{Baudino2015}.
 
 In this paper, we provide the empirical basis for such future investigations by presenting the most comprehensive spectrum of the $\beta$~Pic~b planet to-date. Our data, obtained with the Gemini Planet Imager between 2014 and 2016, covers the {\it Y}, {\it J}, {\it H}, and {\it K} bands. In Section~\ref{sec:obs_data_reduction}, we discuss the observations, data reduction, and spectral extraction. In Section~\ref{sec:Comparison_w_field_ojb}, we compare the spectrum of $\beta$~Pic~b to those of a wide array of brown dwarfs, and present the best fitting objects along with comparisons to low-surface gravity brown dwarf spectral standards. An analysis of the spectrum, along with existing photometry, and comparison to existing models is presented in Section~\ref{sec:results}. Finally, conclusions are discussed in Section~\ref{sec:Discussion}.

\section{Observations and Data Reduction}
\label{sec:obs_data_reduction}
\begin{figure*}
\epsscale{1.2}
\plotone{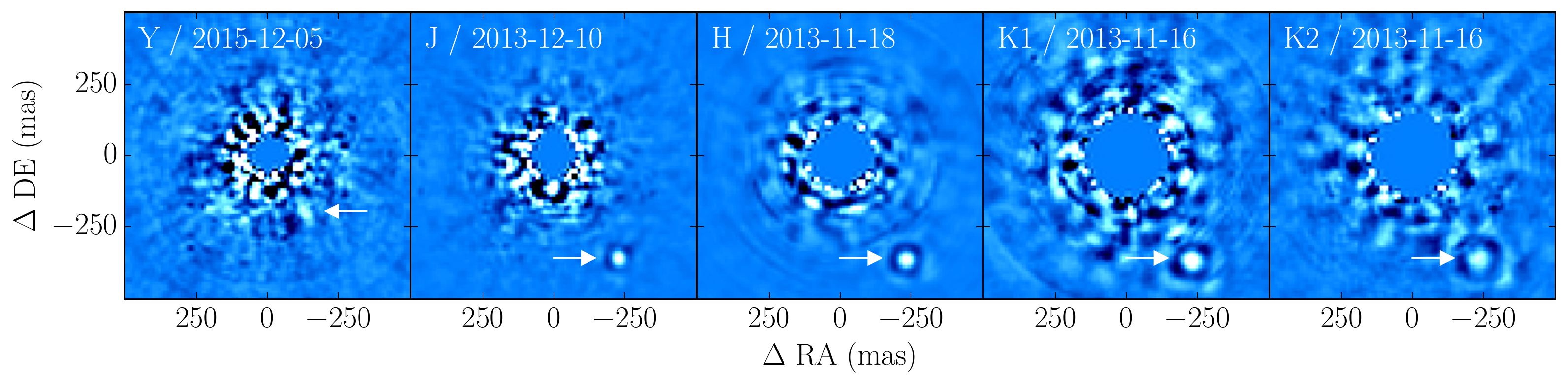}
\caption{PSF-subtracted images of $\beta$~Pic in each of the five GPI filters, with the location of $\beta$~Pic~b highlighted. The images have been rotated such that North is up and East is to the left, with a linear color scale in units of contrast between $\pm2.5\times10^{-5}$ ($\pm11.5$~mags). The significant orbital motion of the planet between 2013 and 2015 is apparent \citep{wang2016}. Each data set was processed using the same KLIP parameters (seven annuli, four segments per annulus, one pixel minimum movement criteria), with a reference PSF constructed from the first ten KL modes. The final images were created by averaging these PSF-subtracted data cubes along the wavelength axis.\label{fig:image_gallery}}
\end{figure*}
\begin{figure*}
\epsscale{1.17}
\plotone{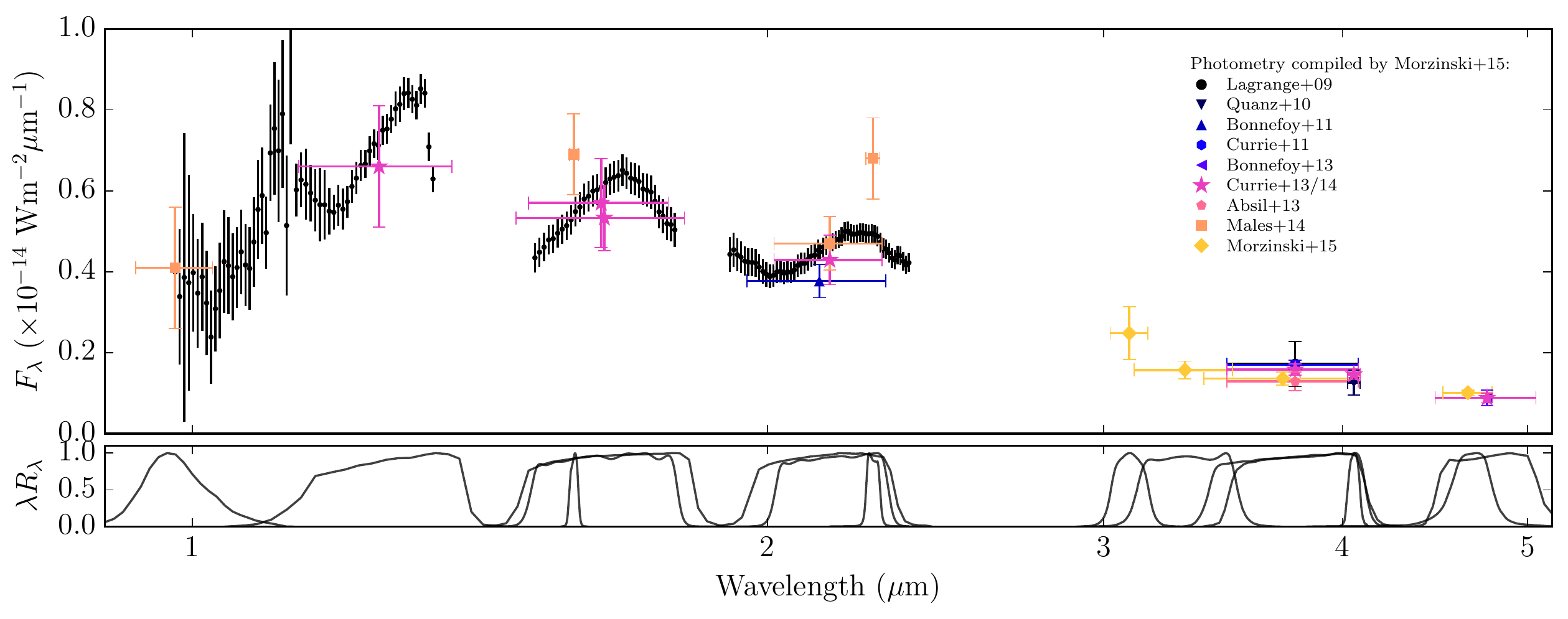}
\caption{(top panel): The GPI spectrum of $\beta$~Pic~b extending from the {\it Y} band to {\it K} band (black points). Photometric measurements of $\beta$~Pic~b, as compiled by \citet{Morzinski2015}, are also plotted (color and symbols given in legend, \citealp{Lagrange2009,Quanz2010,Bonnefoy2011,Currie2011b,Bonnefoy2013,Currie2013,Absil2013,Males2014,Morzinski2015}). (bottom panel): Normalized filter transmission curves for the various photometric measurements of $\beta$~Pic~b.}
\label{fig:SED}
\end{figure*}

The Gemini Planet Imager (GPI) was designed and built to directly image and spectroscopically characterize young, Jupiter-sized, self-luminous extrasolar planets \citep{Macintosh2006,Graham2007}. Installed at Gemini South in the Fall of 2013, GPI underwent commissioning from the Fall of 2013 to the Fall of 2014 before becoming part of the standard instrument suite at Gemini South.

$\beta$~Pic was observed by the GPI Verification and Commissioning team on 2013 November 16 and 18, 2013 December 10 and 11, and 2014 March 23. A log of the observations is given in Table~\ref{tbl:observations}. Observations performed during the instrument commissioning period (November 2013 - November 2014) were not all taken in a stable science environment, and various operational modes were used during a specific data set to evaluate performance of the instrument. For instance, during the 2013 November 18 observations, 32 individual 59.6\,s images were obtained in coronagraphic mode, with the cryocoolers operating at a reduced power level to reduce the effects of vibration introduced into the telescope \citep{Chilcote2012,Larkin2014}.
Observations taken for testing purposes including changes in the AO performance parameters and the vibrations levels of the IFS cryocoolers affect the shape and stability of the GPI point spread function (PSF) on a shorter time scale than would be expected from typical stable operations of Gemini.

{\it Y}-band data were obtained as part a Gemini Large and Long Program focused on the study of debris disks with GPI (GS-2015B-LP-6). These observations occurred when the planet had already moved significantly inwards, resulting in a higher level of noise in the estimated spectrum. The average seeing measured using a Differential Image Motion Monitor (DIMM), total exposure times, and instrument configurations of the observations presented in this paper are listed in Table~\ref{tbl:observations}, along with the specifics of the data-sets that were obtained during verification and commissioning.

Each of these data sets was individually and independently reduced using the GPI pipeline, with standard recipes provided by the GPI Data Reduction Pipeline \citep{Perrin2014}. A short arc lamp exposure was taken with each science observation set to account for offsets of the lenslet spectra due to flexure within the IFS. The GPI data reduction pipeline was used to reduce all images by applying dark corrections, fitting and removing vibration-induced microphonics noise \citep{Chilcote2012,Ingraham2014}, removing bad pixels, fitting satellite spot locations \citep{wang2014}, and extracting each microspectra to create a 37-channel spectral cube. For {\it K}-band data sky frames were subtracted, if available, to remove the thermal background.

High-contrast imaging surveys for faint substellar companions typically use Angular Differential Imaging (ADI, \citealp{Marois2006}) and/or Spectral Differential Imaging (SDI, \citealp{Sparks2002}). These PSF subtraction processes often lead to self-subtraction of any resolved faint companions, creating systematic biases in the extracted photometry that need to be corrected for. Previous studies have used forward modeling approaches where a negative version of the PSF is injected into the reduced images prior to PSF subtraction to estimate the flux and position of faint companions detected in PSF-subtracted images \citep[e.g.,][]{Hinkley2013,Oppenheimer2013,Crepp2015}. For this study, we use a different forward modeling approach that analytically models the effect of stellar PSF subtraction on the PSF of the planet to find the best planet spectrum that matches the signal of the planet after stellar PSF subtraction. We use the generalized method KLIP-FM, described in \citet{Pueyo2016}, which combines the Karhunen-Lo\`{e}ve Image Projection algorithm (KLIP, \citealt{Soummer2012}) and forward modeling. \citet{Pueyo2016} demonstrated the effectiveness of KLIP-FM at reducing the systematic biases inherent in ADI/SDI PSF subtraction by injecting and recovering point sources with known spectra into into GPI {\it J}-band $\beta$~Pic data. KLIP-FM has also been used to measure the astrometry of $\beta$~Pic~b with GPI at milliarcsecond precision \citep{wang2016}. The final PSF-subtracted images of $\beta$~Pic~b in each of the five GPI filters are shown in Figure~\ref{fig:image_gallery}.

As described in \citet{wang2016}, we use the four satellite spots in each spectral channel to estimate the PSF of $\beta$~Pic~b. When using such a PSF fitting method, biases on the spectrum can arise due to a mismatch between the full width at half maximum (FWHM) of the planet and model PSFs. Such a mismatch can occur as a result of the preliminary high-pass filter step before the KLIP PSF subtraction, carried out in order to mitigate the impact of the residual atmospheric halo. We calibrated this effect in an {\it ad-hoc} fashion by exploring an increasing sequence of high-pass filtering cutoff frequencies. Typically the signal-to-noise ratio (SNR) of the planet increases with a more aggressive filter that eliminates the residual AO halo, but such filtering schemes create spurious slopes in the spectrum since they affect the morphology of the planet PSF and of the satellite spots differently. Fortunately, in all data sets considered here the planet SNR is high enough so that there exists a large range of high-pass filtering parameters for which the planet spectrum is stable (between 8 and 15\,px, as defined in \citealt{wang2016}). The resulting spectrum that minimizes the residuals at the location of $\beta$~Pic~b was then estimated by forward modelling, normalized by the average satellite spot intensity in each wavelength channel. An 8000\,K, $\log g=4.0$\,[dex] \citep{Gray2006} {\sc BT-NextGen}\footnote{{\tt https://phoenix.ens-lyon.fr/Grids/BT-NextGen/}} model \citep{Allard2012} convolved to the resolution of GPI, was used to approximate the A6V stellar spectrum of $\beta$~Pic~A. This allows the instrumental and telluric features under identical conditions to be estimated for the planet spectrum and then removed. To exclude low SNR data at the edges of the filter band-passes, we trim the $\beta$~Pic~b spectrum to exclude wavelength channels where the filter transmission is below 80\,\%, excluding 27 of the 185 channels of the full spectrum. The average SNR per resolution element of the final trimmed spectrum, plotted in Figure~\ref{fig:SED}, was 3 in {\it Y}, 17 in {\it J}, 15 in {\it H}, 14 in {\it K1}, and 19 in {\it K2}.

\section{Results \& Discussion}
\label{sec:results}

\subsection{Bolometric Luminosity}
\label{sec:bol_lum}
The bolometric luminosity of $\beta$~Pic~b was most recently estimated as $\log L_{\rm bol}/{\mathcal L}_{\odot}=-3.78\pm0.03$\,[dex] by \citet{Morzinski2015}\footnote{We adopt the same convention as \citet{Morzinski2015} where script letters are used to denote nominal Solar and Jovian values as defined by IAU resolutions.}, calculated using GPI $J$- and $H$-band spectroscopy and a re-calibration of optical through thermal-infrared photometry to remove any systematic bias introduced in previous studies. The bolometric luminosity of $\beta$~Pic~b was reassessed using the new spectroscopic measurements presented in Section~\ref{sec:obs_data_reduction}, which significantly improves the sampling of its spectral energy distribution (SED) in the $Y$ (0.98--1.13\,\micron) and $K$ (1.91--2.38\,\micron) bands.

The procedure is similar to that employed by \citet{Morzinski2015}. The GPI spectrum was combined with band-averaged photometry at 3.31\,\micron, 3.34\,\micron, $L^{\prime}$ (3.80\,\micron), 4.10\,\micron, and $M^{\prime}$ (4.72\,\micron), the values for which are given by \citet{Morzinski2015}. The $Y_{\rm S}$ (0.985\,\micron) and $K$ (2.27\,\micron) photometry measurements used by \citet{Morzinski2015} were rejected from this analysis due to the significant overlap with the new GPI spectrum presented in Figure~\ref{fig:SED}. The measured SED (0.98--4.72\,\micron) was extended to shorter and longer wavelengths using two blackbody functions. The optical blackbody (0.01--0.97\,\micron) was normalized to the integrated flux of the GPI $Y$-band spectrum, while the infrared blackbody (4.73--1000\,\micron) was normalized to the $M^{\prime}$ photometric point. The bolometric luminosity was then measured by integrating the synthetic spectrum formed by the combination of the measured SED and the two blackbody functions. 

The final luminosity and its uncertainty were estimated using a Monte Carlo approach by repeating the integration $10^5$ times. Random draws were made for each trial from each of the band-average photometric points. The individual GPI spectra were varied by drawing from a normal distribution created from a quadratic sum of the band-averaged uncertainty and the satellite spot ratio uncertainty. Conservatively, each point within an individual GPI spectrum was adjusted by the same amount to account for correlation between the spectral channels. The temperature of the two blackbody functions was drawn from a uniform distribution between 1500 and 1900\,K for each trial, and were normalized as described previously. Despite the large range of temperatures, there was no correlation between the choice of temperature for the blackbody extensions and the resulting luminosity. Using the median and 1\,$\sigma$ range of the $10^5$ trials, the bolometric luminosity of $\beta$~Pic~b was found to be $\log L_{\rm bol}/{\mathcal L}_{\odot}=-3.76\pm0.02$\,[dex], consistent with the value reported in \citet{Morzinski2015}. While the choice of a blackbody function is a simplistic one, it only has a small contribution to the total flux of $\beta$~Pic~b, with the short- and long-wavelength blackbody extensions contributing $3\pm1$\,\% and $14\pm1$\,\%, respectively.

\begin{figure}
\epsscale{1.2}
\plotone{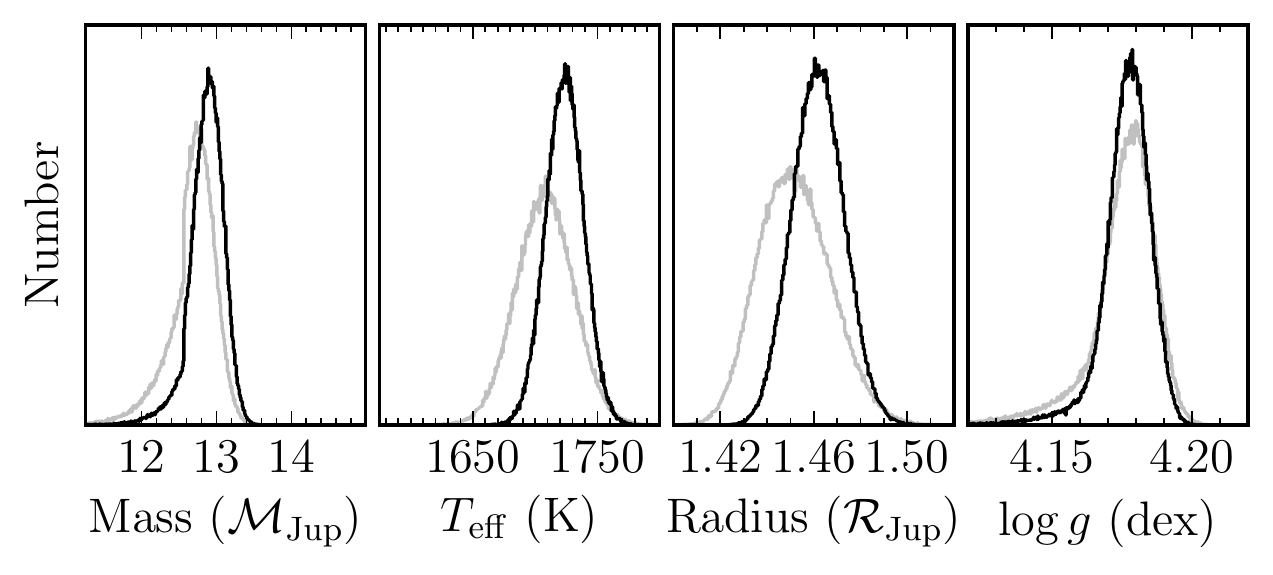}
\caption{Histograms of the values of the four parameters from the Monte Carlo analysis comparing the empirical luminosity and age of $\beta$~Pic~b to the \citet{2003A&A...402..701B} hot-start evolutionary models. This analysis was performed using the luminosity and age presented in this paper (black histogram), and using the values prented in \citet{Morzinski2015} ($\log L_{\rm bol}/{\mathcal L}_{\odot} = -3.78\pm0.03$\,[dex] and $t = 23\pm3$~Myr; gray histogram). The asymmetric distribution for mass is caused by the significant increase in the predicted luminosity due to the onset of deuterium burning.}
\label{fig:params}
\end{figure}

The bolometric luminosity of $\beta$~Pic~b and the age estimate for the system of $24\pm3$~Myr \citep{Bell2015} were compared to the \citet{2003A&A...402..701B} hot-start evolutionary models\footnote{{\tt https://phoenix.ens-lyon.fr/Grids/AMES-Cond/ISOCHRONES/}} to derive a model-dependent estimate of the mass ($M$), temperature ($T_{\rm eff}$), radius ($R$), and surface gravity ($g$) of $\beta$~Pic~b. A Monte Carlo procedure was used to propagate the uncertainty of the luminosity ($L_{\rm bol}$) and age ($t$) to the four derived parameters. At each step, a random luminosity and age were drawn from two normal distributions, one in $\log L_{\rm bol}$ and the other in $t$. The model grid was linearly interpolated first in $\log t$ to the randomly selected age, and then in $\log M$ to an arbitrarily high resolution. Interpolation was performed in $\log L_{\rm bol}$, $\log T_{\rm eff}$, $R$, and $\log g$ due to their behavior as a function of $\log t$ and $\log M$. The randomly selected luminosity was then used to select a model within the interpolated grid. This process was repeated $10^5$ times yielding $M=12.9\pm0.2$\,\mjup, $T_{\rm eff} = 1724\pm15$\,K, $R = 1.46\pm0.01$\,\rjup, and $\log g = 4.18\pm0.01$\,[dex] (Figure~\ref{fig:params}). These are consistent consistent with the results of a previous analysis by \citet{Morzinski2015}, who reported a mass, effective temperature, and radius for $\beta$~Pic~b of $12.7\pm0.3$\,\mjup, $1708\pm23$\,K, and $1.45\pm0.02$\,\rjup, respectively.

\subsection{Comparison with field objects}
\label{sec:Comparison_w_field_ojb}
The spectrum of $\beta$~Pic~b was compared with a library of 1600 M-, L-, and T-dwarf spectra compiled from the SpeX Prism library\footnote{{\tt http://pono.ucsd.edu/\~{}adam/browndwarfs/spexprism}} \citep{2014ASInC..11....7B}, the IRTF Spectral Library\footnote{{\tt http://irtfweb.ifa.hawaii.edu/\~{}spex/IRTF\_Spectral\_Library}} \citep{2005ApJ...623.1115C}, the Montreal Spectral Library\footnote{{\tt https://jgagneastro.wordpress.com/the-montreal-spectral-library/}} (e.g., \citealp{Gagne:2015dc,2016arXiv160706117R}), and the sample of young ultracool dwarfs presented in \citet{Allers2013}. The spectral types for the objects within the library were obtained from a number of literature sources, and are given for the individual objects described later in this section. The near-infrared spectral type was used for objects with both an optical and near-infrared spectral type. The literature was also searched to obtain the surface gravity classifications for each object, using either of the schemes outlined by \citet{2005ARA&A..43..195K, 2006ApJ...639.1120K,2009AJ....137.3345C} ($\alpha$, $\beta$, $\gamma$, $\delta$, in descending order of surface gravity), or \citet{Allers2013} ({\sc fld-g}, {\sc int-g}, {\sc vl-g}, similarly). Briefly, both classification schemes categorize ultracool dwarfs into three groupings: field surface gravity consistent with that seen for old field dwarfs ($\alpha$, {\sc fld-g}), intermediate surface gravity ($\beta$, {\sc int-g}), and very low surface gravity consistent with that seen for young brown dwarfs ($\gamma$, {\sc vl-g}). \citet{2005ARA&A..43..195K} define a fourth classification, $\delta$, for objects which exhibit even stronger low-gravity features in their spectra than those classified as $\gamma$.

The spectrum of each object was degraded to the spectral resolution of GPI (between $\lambda/\delta\lambda=35$ at $Y$ and $\lambda/\delta\lambda=79$ at $K2$) by convolution with a Gaussian function of the appropriate width. The uncertainties were similarly degraded, normalized by the effective number of spectral channels within the convolution window. The spectrum of $\beta$~Pic~b was compared to this library using three different procedures. First, the five GPI bands were fit independently to explore the sensitivity of each bandpass to the spectral type and surface gravity of $\beta$~Pic~b. Second, the five bands were fit simultaneously but were each normalized independently and without constraint to account for both the dispersion in near-infrared colors of young low-gravity brown dwarfs (e.g., \citealp{Leggett2003}), and for the uncertainty in the photometric calibration of the GPI data. Third, the five bands were fit simultaneously as before, but the normalization of each band was restricted by the uncertainty of the photometric calibration of the GPI data \citep{2014SPIE.9147E..85M}. We find that all three methods yield similar results in terms of the spectral type of $\beta$~Pic~b, and provide strong evidence for a low surface gravity.

\subsubsection{Fits to the individual bands}
\begin{figure}
\epsscale{1.2}
\plotone{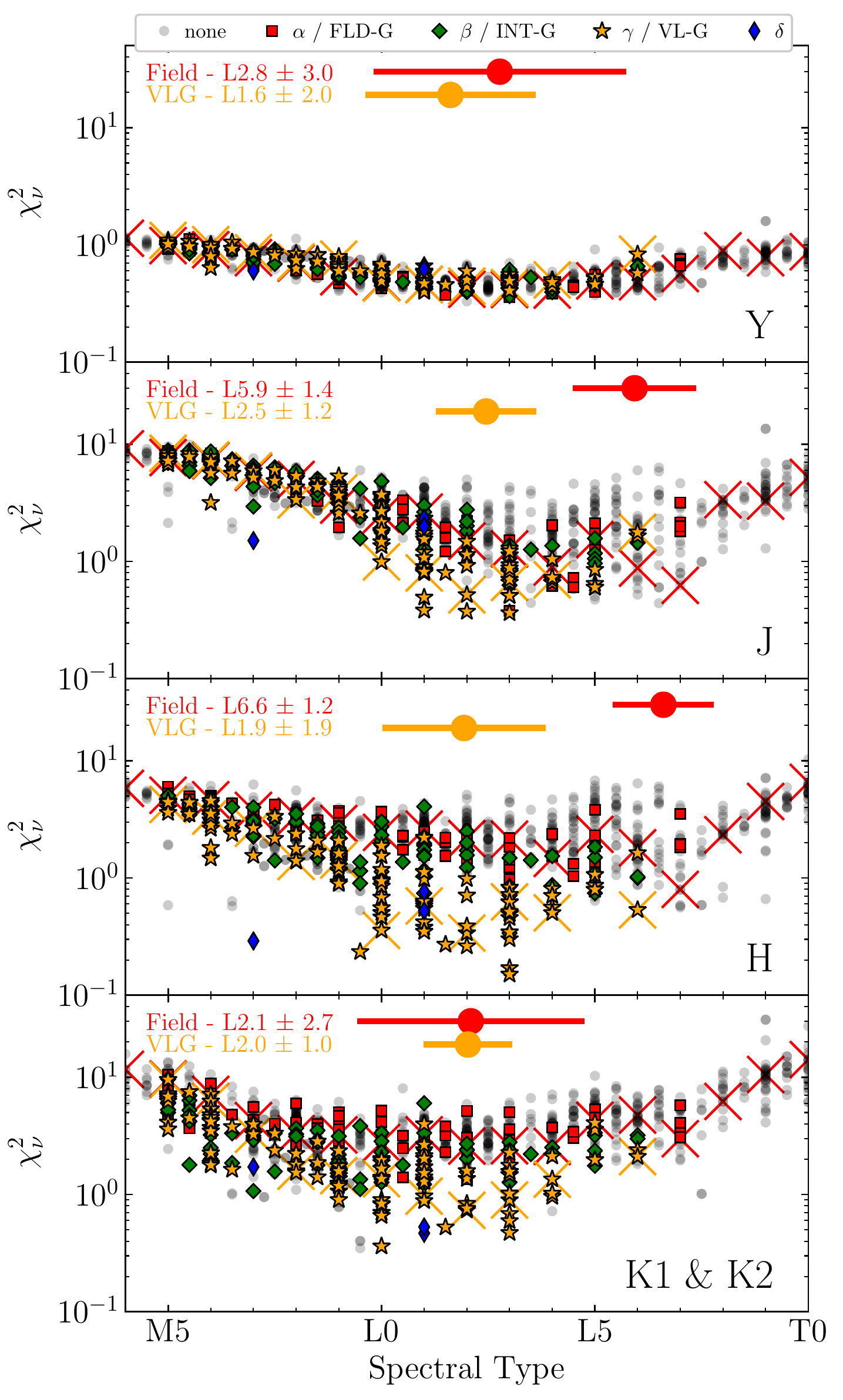}
\caption{$\chi^2_{\nu}$ as a function of spectral type for the M, L and T dwarfs within the spectral library fit to the spectum of $\beta$~Pic~b in each of the GPI bandpasses. The {\it K1} and {\it K2} spectra were combined to create a single {\it K}-band spectrum. Different markers were used to indicate the different gravity classes using the scheme described in \citet{Allers2013}, with the legend given at the top of the figure. The optical and near infrared gravity classification have been grouped together for clarity. Objects without any gravity classification are plotted as gray circles. Spectral standards for field-gravity \citep{Burgasser:2006cf,Kirkpatrick:2010dc} and low-gravity \citep{Allers2013} objects are highlighted with large red and yellow crosses, respectively.}
\label{chi2a}
\end{figure}
The spectrum of $\beta$~Pic~b in each of the four near-infrared bands ({\it YJHK}) was fit to the corresponding spectrum of each comparison object within the library. The {\it K}-band spectrum of $\beta$~Pic~b was created by combining the GPI {\it K1} and {\it K2} spectra, discarding the overlapping spectral channels within the {\it K1} spectrum due to systematics in the {\it K1} spectrum. The spectrum of the comparison object was multiplied by a scaling factor to account for the different distance and radius between that object and $\beta$~Pic~b. The optimal scaling factor was found analytically for each object and band \citep[e.g.,][]{2016ApJ...820...32B}. The uncertainty on the spectrum of $\beta$~Pic~b and the comparison object were added in quadrature. The number of degrees of freedom was typically 29 for {\it Y} band, 32 for {\it J} band, 34 for {\it H} band, and 56 for {\it K} band. The minimum $\chi^2_{\nu}$ for each object in each band is plotted as a function of spectral type in Figure~\ref{chi2a}.

The sensitivity of the {\it J}-, {\it H}-, and {\it K}-band spectra to surface gravity is apparent in Figure~\ref{chi2a}, with the low-surface gravity objects typically providing a better fit to $\beta$~Pic~b than field-gravity objects of the same spectral type. Given the low resolution of the GPI data, this sensitivity is primarily due to differences in the shape of the continuum between field and low-gravity objects \citep{Allers2013}, rather than differences in the strengths of gravity-sensitive absorption lines. The difference between the spectra of field and low-gravity objects is most pronounced in the {\it H} and {\it K} bands, where the minimum $\chi^2_{\nu}$ of the low-surface gravity objects is significantly lower than that of field-gravity objects of the same spectral type (Fig.~\ref{chi2a}).

We estimated the spectral type of $\beta$~Pic~b in each band by a comparison to the spectra of field surface gravity standards \citep{Burgasser:2006cf,Kirkpatrick:2010dc}, and low-surface gravity standards \citep{Allers2013}. The weighted average of the numerical spectral types of the standards, weighted according to the ratio of their $\chi^2$ to the minimum $\chi^2$ of all the standards (e.g., \citealp{2010ApJ...710.1142B}), was adopted as the spectral type. A systematic uncertainty of one half subtype was assumed for the standards. This process was repeated for both surface gravity subsets, and for each of the five bands. The adopted spectral type and corresponding uncertainty for $\beta$~Pic~b are given for each band in Figure~\ref{chi2a}, ranging from L2 to L6.5 for the field surface gravity standards and from L1.5 to L2.5 for the low-surface gravity standards, both rounded to the nearest half subtype.

\subsubsection{Unrestricted fit to the full spectrum}
\label{sec:unrestricted}
\begin{figure}
\epsscale{1.15}
\plotone{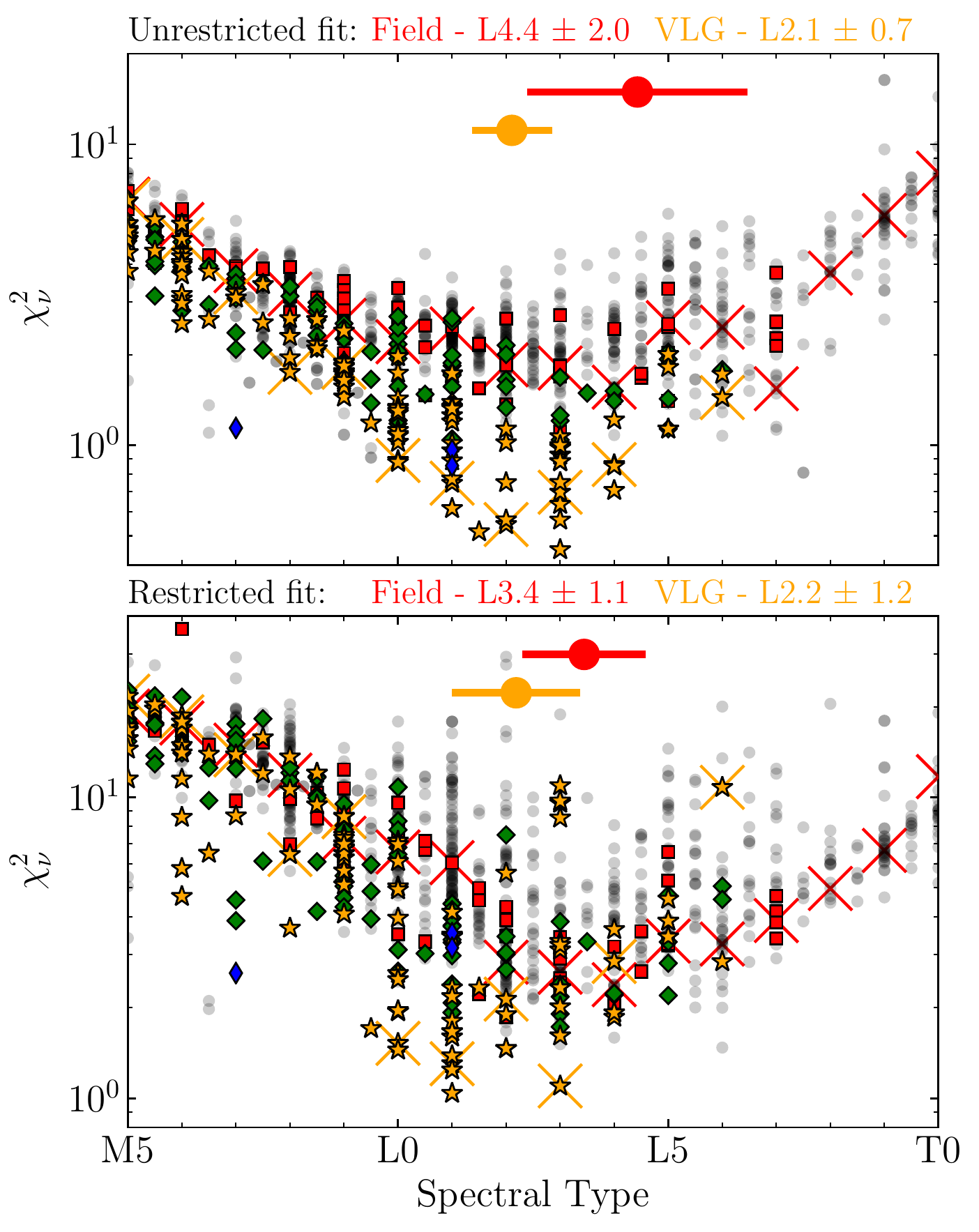}
\caption{$\chi^2_{\nu}$ as a function of spectral type for each object within the spectral library for the unrestricted fit described in \S~\ref{sec:unrestricted} (top panel), and for the restricted fit described in \S~\ref{sec:restricted} (bottom panel), to the GPI spectrum of $\beta$~Pic~b. The symbols are as in Figure~\ref{chi2a}. In both cases, the low-gravity objects typically have lower $\chi^2_{\nu}$ values than field-gravity objects of the same spectral type. The spectral type of $\beta$~Pic~b was estimated for both gravity subsets, with the estimates being consistent between the two different fitting procedures.}
\label{chi2b}
\end{figure}
The full GPI spectrum of $\beta$~Pic~b was then fit to each object within the library. Each band of the spectrum was scaled independently to account for the dispersion in near-infrared colors seen for brown dwarfs of a given spectral type \citep{Leggett2003}, and for the uncertainty in the absolute flux calibration of the GPI data \citep{2014SPIE.9147E..85M}. This was achieved by summing the $\chi^2$ of each object in each band, equivalent to fitting the five bands simultaneously with five independent scale factors. The resulting minimum $\chi^2_{\nu}$ values for each object are plotted in Figure~\ref{chi2b} (top panel). The number of degrees of freedom was typically 152, but was lower for objects that had limited spectral coverage. As with the previous fit, the spectral type of $\beta$~Pic~b was estimated as L$4\pm 2.5$ using the field-gravity standards, and L$2\pm 1$ using the low-gravity standards, consistent with previous estimates based on fits to the broadband photometry of $\beta$~Pic~b \citep{Males2014}. The significantly lower $\chi^2_{\nu}$ values for the low-gravity objects within the library provides strong evidence for the low surface gravity of $\beta$~Pic~b, consistent with previous photometric and spectroscopic analyses \citep{Chilcote2015,Morzinski2015}.

\begin{figure*}
\epsscale{1.15}
\plotone{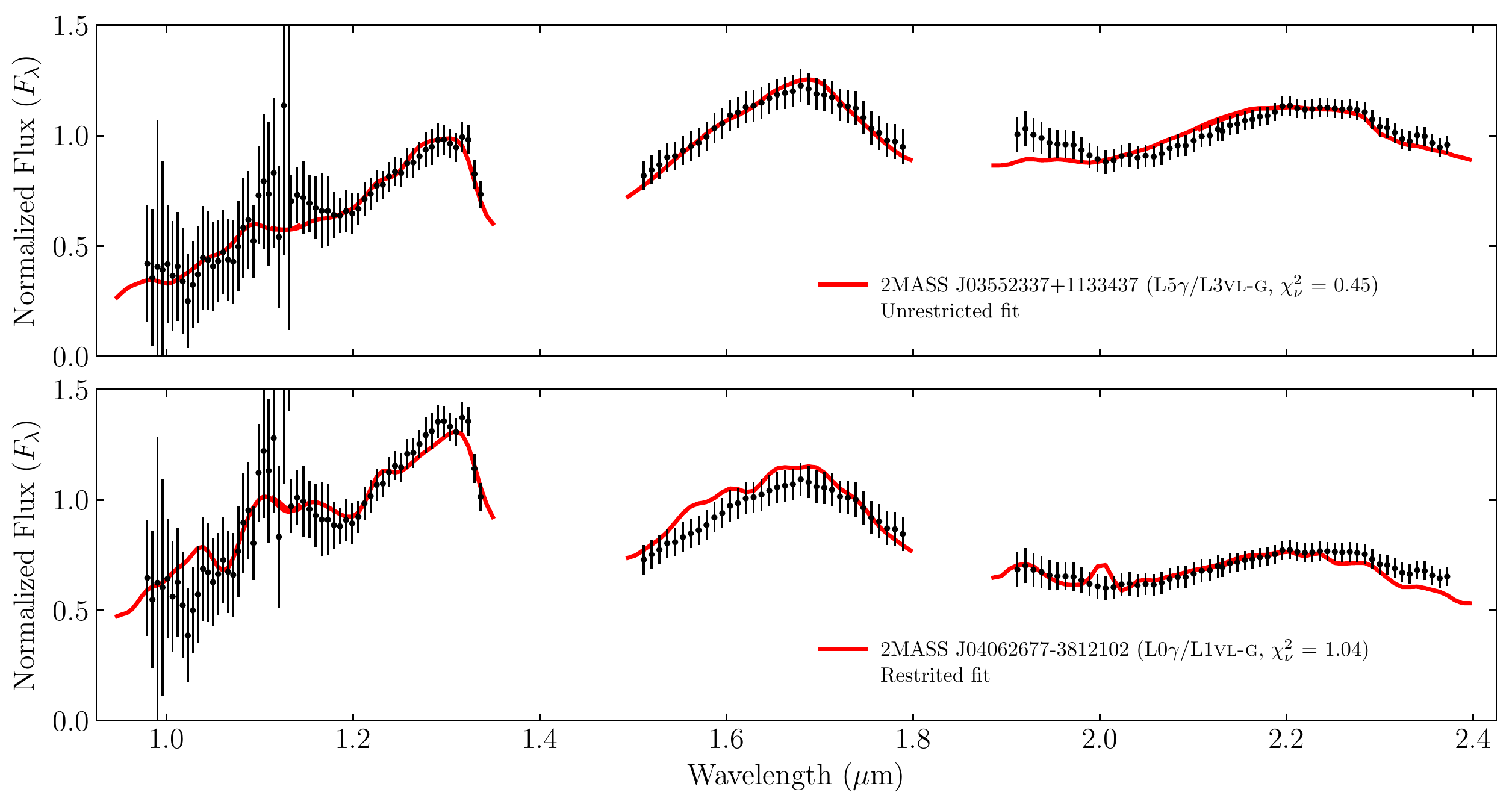}
\caption{The best fit object to the spectrum of $\beta$~Pic~b within the spectral library for the unrestricted (top panel, \citealp{Allers2013}) and restricted (bottom panel, \citealp{Kirkpatrick:2010dc}) fits. The optical and near-infrared spectral type and gravity classifications are given for both objects.}
\label{two_panel_flux_vs_wavelength}
\end{figure*}

Of all the objects within the library, the best fit object from the unrestricted fit was found to be \objectname[2MASS J03552337+1133437]{2MASS~J03552337+1133437} (2M~0355+11, $\chi^2_{\nu} = 0.45$), a nearby (8--9~pc, \citealp{Faherty:2013bc,Liu:2013ej}) and extremely red \citep{2009AJ....137.3345C} brown dwarf with a near-infrared (optical) spectral type of L3 {\sc vl-g} (L5$\gamma$). The spectrum of 2M~0355+11 is plotted with $\beta$~Pic~b in Figure~\ref{two_panel_flux_vs_wavelength} (top panel). Based on a kinematic analysis and the spectral signatures of youth, 2M~0355+11 is a confirmed member of the $149^{+51}_{-19}$~Myr \citep{Bell2015} AB Doradus moving group \citep{Faherty:2016fx}. Due to the unusual near-infrared spectrum of 2M~0355+11, \citet{Gagne:2015dc} assign it a special spectral classification of L3--L6$\gamma$, and classify objects with similar spectra as {\it J0355-type}. These objects are visually similar to L4$\gamma$ objects but with a shallower CO band at 2.3\,\micron\ \citep{Gagne:2015dc}. The spectrum of 2M~0355+11 exhibits strong indicators of low surface gravity, and the unusual near- and mid-infrared colors were explained by flux redistribution to longer wavelengths due to enhanced dust or thick clouds in the photosphere \citep{Faherty:2013bc, Faherty:2016fx}.

A good fit was also found to the spectrum of \objectname[2MASS J22351658-3844154]{2MASS~J22351658--3844154} ($\chi^2_{\nu}=0.52$), an L1.5$\gamma$ candidate member of the $45\pm4$~Myr \citep{Bell2015} Tucana-Horologium moving group \citep{Gagne:2015dc}. Of the low-gravity near-infrared standards defined by \citet{Allers2013}, the best fit was found to be the L2 {\sc vl-g} standard \objectname[2MASSI J0536199-192039]{2MASSI~J0536199--192039} ($\chi^2_{\nu} = 0.55$), a candidate member of both the $42^{+6}_{-4}$\,Myr \citep{Bell2015} Columba moving group \citep{Gagne:2014gp,Gagne:2015dc}, and the $24\pm3$\,Myr \citep{Bell2015} $\beta$~Pictoris moving group \citep{Faherty:2016fx}. This object, and the other early- to mid-L near-infrared spectral standards from \citet{Allers2013} are plotted in Figure~\ref{fig:vlg_std}.

\subsubsection{Restricted fit to the full spectrum}
\label{sec:restricted}
Finally, we fit the morphology within each band and the relative flux levels of the different bands by restricting the range over which the scale factor for each band can vary based on the expected photometric accuracy of GPI. In order to restrict this range, the $\chi^2$ equation was modified with a cost term based on a comparison of the scale factor for a band, and the uncertainty on the satellite spot ratio in that band \citep{2014SPIE.9147E..85M}. The $\chi^2$ for the $k^{\rm th}$ comparison object was calculated as:

\begin{multline}\label{eq:1}
    \chi^2_k = \sum_{j=0}^4 \sum_{i=0}^{n_j} \left[\frac{F_j(\lambda_i) - \alpha_k\beta_{j,k} C_{j,k}(\lambda_i)}{\sqrt{\sigma^2_{F_j}(\lambda_i) + \sigma^2_{C_{j,k}}(\lambda_i)}}\right]^2 \\
    + \sum_{j=0}^4 n_j\left[\frac{\beta_{j,k}-1}{\sigma_{m_j}}\right]^2
\end{multline}
where $F_j(\lambda_i)$ and $\sigma_{F_j}(\lambda_i)$ are the flux and uncertainty of $\beta$~Pic~b in the $i^{\rm th}$ wavelength channel of the $j^{\rm th}$ band and $C_{j, k}(\lambda_i)$ and $\sigma_{C_{j,k}}(\lambda_i)$ is the flux and uncertainty of the $k^{\rm th}$ comparison object in the same channel and band. The spectrum of the comparison object is multiplied both by a scale factor $\alpha_k$ which is the same for each band, and by an additional scale factor for the $j^{\rm th}$ band $\beta_{j,k}$. The first term of Equation~\eqref{eq:1} gives the standard $\chi^2$ equation, summed over all $n_j$ wavelength channels in the $j^{\rm th}$ band, and over all five bands. This is modified by a cost term which compares the band-dependent scaling factor $\beta_{j,k}$ to the fractional uncertainty of the satellite spot flux ratio $\sigma_{m_j}$ for the $j^{\rm th}$ band \citep{2014SPIE.9147E..85M}.

The minimum $\chi^2_{\nu}$ for each object is plotted for this restricted fit in Figure~\ref{chi2b} (bottom panel). The number of degrees of freedom was typically 151, but was lower for objects that had limited spectral coverage. As with the two previous fits, the spectral type of $\beta$~Pic~b was estimated using the field-gravity and low-gravity standards as L$3.5\pm1.0$ and L$1.5\pm1.5$, respectively. While these estimates of the spectral type are consistent with the results of the unrestricted fit, the minimum $\chi^2_{\nu}$ values are higher for each object due to the additional cost term included in the restricted fit, an effect that is most pronounced for the mid to late-type M-dwarfs within the library.

\begin{figure*}
\epsscale{1.15}
\plotone{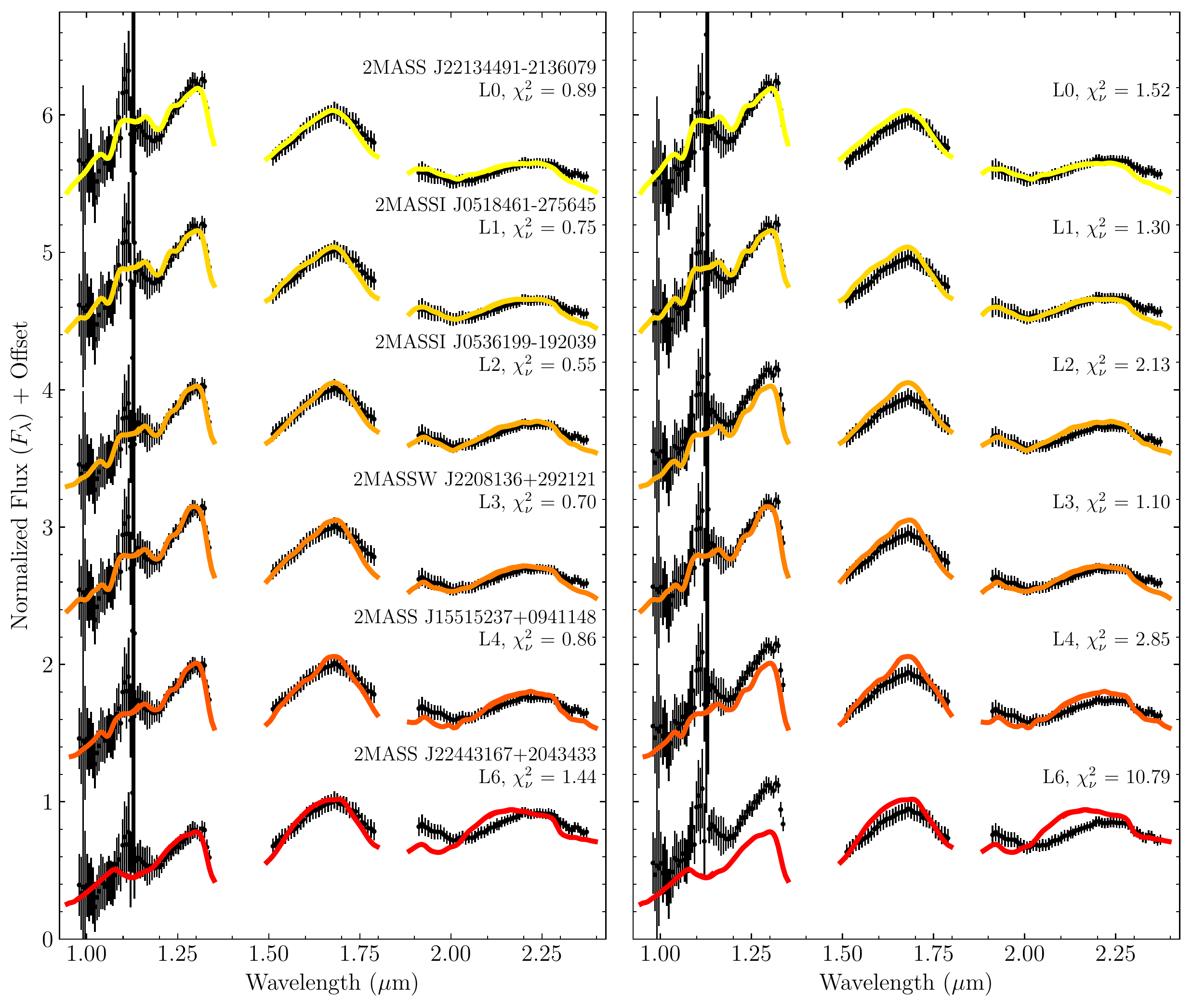}
\caption{The spectrum of $\beta$~Pic~b (black points) fit to the L-dwarf near-infrared low-gravity standards from \citet{Allers2013} (solid curves) using the unrestricted fit described in Section~\ref{sec:unrestricted} (left panel) and the restricted fit described in Section~\ref{sec:restricted} (right panel). The spectrum of each standard is normalized to the flux at 1.65\,\micron, and then offset for clarity}.
\label{fig:vlg_std}
\end{figure*}

For the restricted fit case, the best fitting result from the spectral library is \objectname[2MASS J04062677-381210]{2MASS~J04062677--381210} (2M~0406--38, $\chi^2_{\nu}=1.04$), a brown dwarf with an L0$\gamma$/L1 {\sc vl-g} (optical/near-infrared, \citealp{Faherty:2013bc, Allers2013}) spectral type (Figure~\ref{two_panel_flux_vs_wavelength}, bottom panel). The kinematics of 2M~0406--38 are ambiguous in terms of nearby moving group membership, being a probable member of several nearby moving groups as well as having consistent space motion as old field objects \citep{Faherty:2016fx}. Other objects with a good fit include \objectname[2MASS J01415823-4633574]{2MASS~J01415823--4633574} ($\chi^2_{\nu} = 1.46$); an L0$\gamma$/L2 (optical/near-infrared, \citealp{2009AJ....137.3345C, 2014AJ....147...34S}) high-probability candidate member of the Tucana-Horologium moving group \citep{Gagne:2014gp}; and the \citet{Allers2013} near-infrared low-gravity standards \objectname[2MASSW J2208136+292121]{2MASSW~J2208136+292121} (L3 {\sc vl-g}, $\chi^2_{\nu} = 1.10$), a candidate member of the $\beta$~Pictoris moving group \citep{Liu:2016co}, and \objectname[2MASSI J0518461-275645]{2MASSI~J0518461--275645} (L1 {\sc vl-g}, $\chi^2_{\nu} = 1.30$), a probable member of the Columba moving group \citep{Liu:2016co}. All of the \citet{Allers2013} low-gravity standards are shown in Figure~\ref{fig:vlg_std} (right panel).

\subsubsection{Spectral type and gravity classification of $\beta$~Pic~b}
Based on a comparison of the full GPI spectrum of $\beta$~Pic~b to the \citet{Allers2013} low-gravity standards, the spectral type of $\beta$~Pic~b was estimated using the unrestricted and restricted procedures as L$2.1\pm0.7$ and L$2.2\pm1.2$, respectively. While these two estimates are consistent with one another for $\beta$~Pic~b, we would expect the unrestricted fit to more reliably estimate the spectral type of young low-gravity objects due to the observed range of their near-infrared colors. Rounding to the nearest half subtype, we adopt a spectral type of L$2\pm1$ for $\beta$~Pic~b, consistent with previous photometric and spectroscopic estimates of L2--5 \citep{Currie2013}, L$2\gamma\pm2$ \citep{Bonnefoy2013}, L$2.5\pm1.5$ \citep{Males2014}, and L$1^{+1}_{-1.5}$ \citep{Bonnefoy2014}. The significantly improved fits to the low-gravity objects within the spectral library (Figures~\ref{chi2a} and \ref{chi2b}) demonstrates that $\beta$~Pic~b has a near-infrared spectrum consistent with that of a low-surface gravity object, and as such we assign it a surface gravity classification of $\gamma$. We do not assign a classification in the \citet{Allers2013} scheme as the bandwidths of the indices used to define this scheme are smaller than the spectral resolution of the GPI spectrum.

This spectral type estimate was converted into a bolometric luminosity using the $J$- and $K_S$-band empirical spectral type to bolometric correction relations for young, low-gravity objects derived by \citet{2015ApJ...810..158F}. We estimate an absolute $J$-band magnitude in the MKO system \citep{2002PASP..114..180T} of $\beta$~Pic~b from the flux-calibrated GPI spectrum of $M_J=12.56 \pm 0.08$, a bolometric correction of $BC_J = 1.48\pm0.28$, a bolometric magnitude of $M_{\rm bol}=14.04\pm0.29$, and a bolometric luminosity of $\log L_{\rm bol}/{\mathcal L}_{\odot} = -3.72\pm0.12$\,[dex]. Similarly, for $K_S$-band: $M_K = 10.86\pm0.15$, $BC_K = 3.26\pm0.13$, $M_{\rm bol}=14.11\pm0.20$, $\log L_{\rm bol}/{\mathcal L}_{\odot} = -3.75\pm0.08$\,[dex]. Both of these luminosity estimates are consistent with the empirical bolometric luminosity of $\beta$~Pic~b presented in Section~\ref{sec:bol_lum}. We also convert the absolute $H$-band magnitude of $\beta$~Pic~b ($M_H = 11.80\pm0.09$) into an effective temperature of $1681\pm64$\,K using the relations derived by \citet{2015ApJ...810..158F}. The spectral type was also converted into an effective temperature using the relations presented in \citet{Faherty:2016fx}. Using the polynomial fit to {\it bona fide} and high-likelihood moving group members, the spectral type of $\beta$~Pic~b corresponds to an effective temperature of $1847\pm242$\,K. Including probable moving group members in the polynomial fit increases the derived effective temperature to $1888\pm215$\,K, while including both probable moving group members and T-dwarf imaged planetary-mass companions decreases it to $1787\pm240$\,K. These estimates are consistent with the effective temperature estimated from the evolutionary models in Section~\ref{sec:bol_lum}.

\subsection{Comparison with atmospheric models}
\label{sec:atmo_models}
\begin{figure*}
\epsscale{1.17}
\plotone{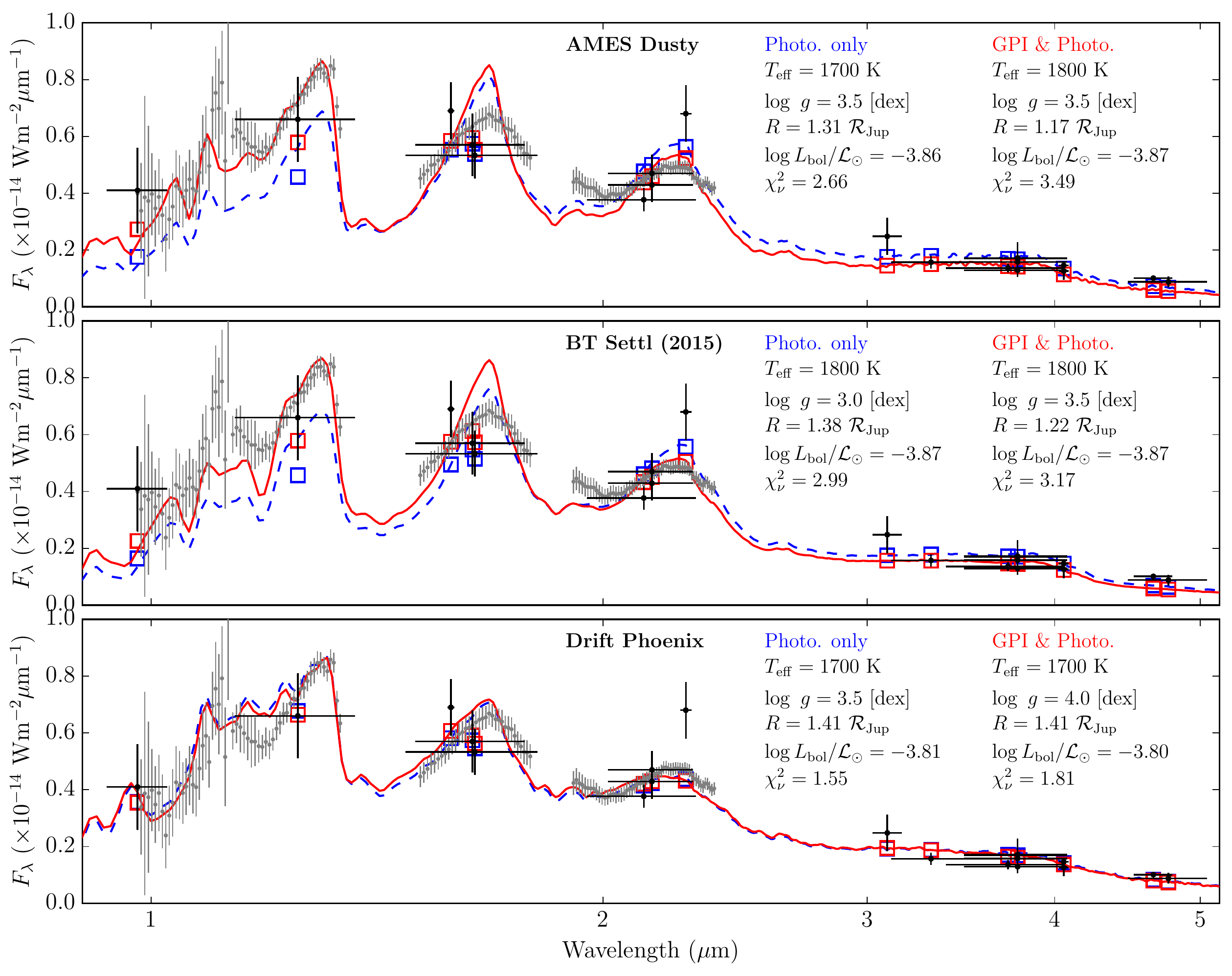}
\caption{The best fit models within each of the three atmospheric model grids found using only the photometric measurements (blue dashed curve) and using both the photometric and spectroscopic measurements (red solid curve) of $\beta$~Pic~b. The photometric measurements of $\beta$~Pic~b compiled by \citet{Morzinski2015} are plotted as black points, while the GPI spectra presented in this study are plotted as light gray points. Synthetic photometry (open blue and red squares) was computed for each model using the filter profiles shown in Figure~\ref{fig:SED}.}
\label{fig:SED_fitted}
\end{figure*}
\begin{figure*}
\epsscale{1.17}
\plotone{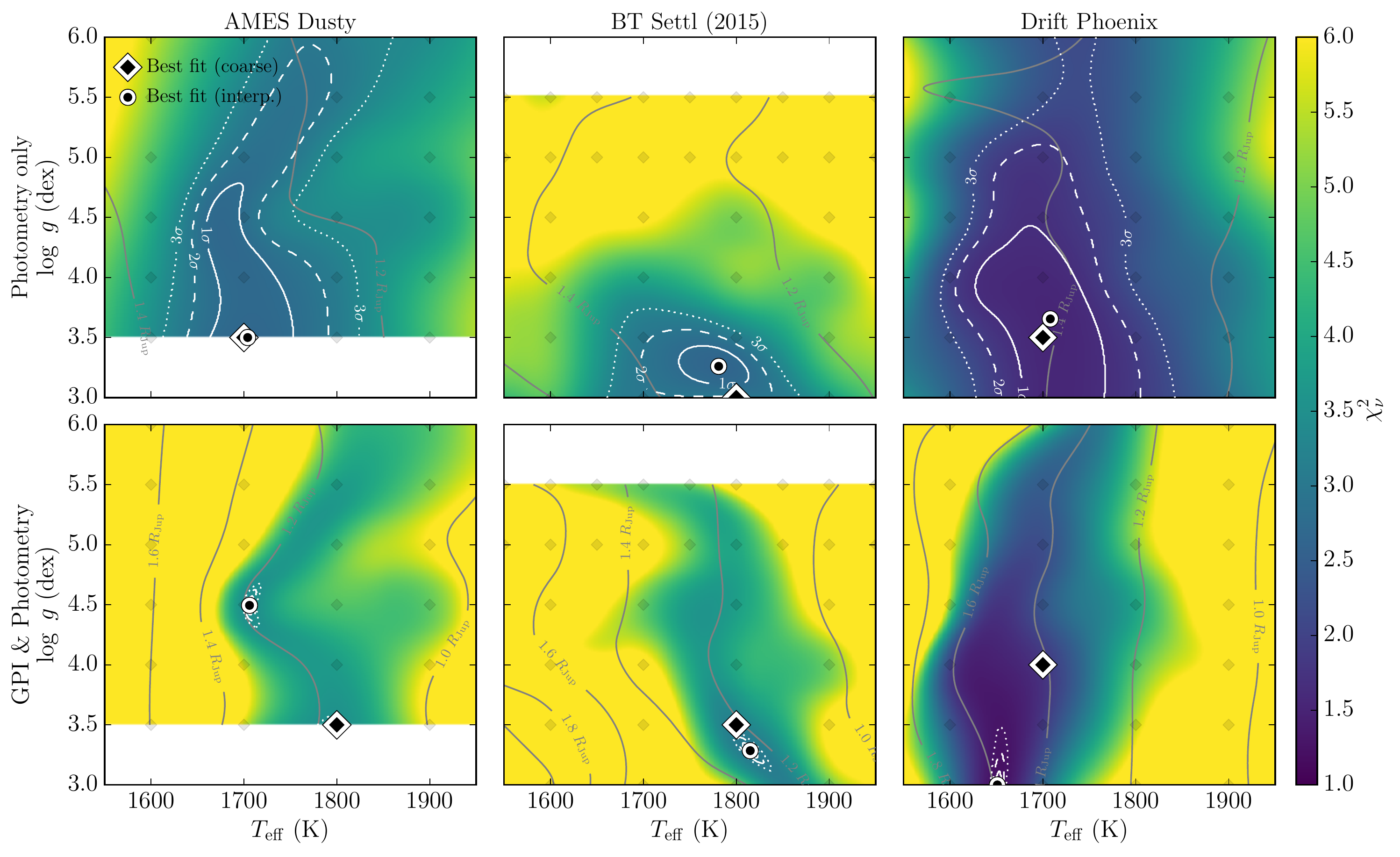}
\caption{Goodness of fit statistic ($\chi^2_{\nu}$) for the {\sc Ames-Dusty}, {\sc BT-Settl} (2015) and {\sc Drift Phoenix} model grids as a function of effective temperature and surface gravity. Grid points are indicated with light-gray diamonds. The points between model grid points were linearlly interpolated version of the grid, with a spacing of 1\,K for $T_{\rm eff}$ and 0.005\,[dex] for $\log g$. The best fit model within the original grid is indicated by a large diamond, with the best fit model within the interpolated grid indicated by a circle. The white contours indicate the 68\,\% (solid), 95\,\% (dashed), and 99\,\% (dotted) confidence intervals, calculated using the $\chi^2$.}
\label{fig:logg_teff}
\end{figure*}

Using the spectral data obtained with GPI, we computed updated best spectral model fits combining GPI spectral data with previously published photometry of $\beta$~Pic~b \citep{Lagrange2009,Quanz2010,Bonnefoy2011,Currie2011b,Bonnefoy2013,Currie2013,Absil2013,Males2014,Morzinski2015}. The SED of $\beta$~Pic~b was compared to publicly available grids of model atmospheres: {\sc Ames-Dusty}\footnote{{\tt https://phoenix.ens-lyon.fr/Grids/AMES-Dusty}} \citep{Chabrier2000,Allard2001}, {\sc BT-Settl} (2015)\footnote{{\tt https://phoenix.ens-lyon.fr/Grids/BT-Settl/CIFIST2011\_2015}} \citep{Allard2012} and {\sc Drift Phoenix}\footnote{{\tt http://svo2.cab.inta-csic.es/theory/newov}} \citep{Helling2008,Woitke2003,Helling2006}. All of these model grids are calculated using the {\sc Phoenix} atmosphere models. The {\sc Ames-Dusty} grid combines the NASA AMES molecular H$_2$O and TiO line lists and includes the treatment for the condensation of dust within the atmosphere. The {\sc BT-Settl} (2015) models are part of the BT model family, using updated line lists and revised solar abundances. {\sc BT-Settl} uses a detailed cloud model to define the distribution of condensates within the atmosphere. The {\sc Drift Phoenix} model grids combine the {\sc Phoenix} model with the non-equilibrium cloud model {\sc Drift}.

The fitting procedure was similar to that described in Section~\ref{sec:Comparison_w_field_ojb} for the individual spectra in each model grid. The spectra were convolved such that their spectral resolution matched the spectral resolution in each of the GPI wavelength bands \citep{Larkin2014}. To compute synthetic photometry, the model spectra were integrated over the bandpass using filter curves published for each individual filter and instrument. The $(\chi^2_{\nu})$ statistic for each model in comparison to the spectral data was calculated using the method described in Section~\ref{sec:Comparison_w_field_ojb} and using Equation~\eqref{eq:1}, where $C_{j, k}(\lambda_i)$ is the flux of model spectrum in the same channel and band. The $\chi^2_{\nu}$ statistic was calculated for each band, and for the spectral bands a punitive factor to account for the uncertainty on the satellite spot ratio in that band was used. We compute this best fit result using only the existing photometry points and for the photometry points combined with the GPI spectrum. The best fit results span a range of grid models from 1700--1800\,K, with a \logg$=3$--$4$\,[dex] and a $R = 1.17$--$1.41$\,\rjup. The best fit to the combined photometry and spectroscopy is found in the {\sc Drift Phoenix} grid (\teff$ = 1700$\,K, \logg$=4.0$\,[dex], $R=1.41$\,\rjup, $\chi^{2}_{\nu} = 1.81$). The best fit spectrum for each of the different models for both photometry only and GPI spectrum and photometry are shown in Figure~\ref{fig:SED_fitted} and the results are shown in Table~\ref{tbl:best_fit_results}.

This process was repeated on an interpolated version of each grid, with the points between grid points interpolated using a quadratic spline in the logarithm of the flux, where the spacing of $T_{\rm eff}$ and $\log g$ were reduced to an arbitrarily high resolution of 1\,K and 0.005\,[dex], respectively. The grids were also interpolated using a bilinear interpolation scheme which produced similar results. We ran the same analysis as above and find that the best fit results span a range of grid models from 1651--1815\,K, with a \logg$= 3.00$--$4.50$\,[dex] and a $R = 1.18$--$1.58$\,\rjup. Again, the best fit grid is produced by the {\sc Drift Phoenix} with \teff\,$ = 1651$\,K, \logg$=3.00$\,[dex], $R=1.58$\,\rjup\ and a $\chi^{2}_{\nu} = 1.21$. The $\chi^2_{\nu}$ surfaces for the interpolated grids are shown in Figure~\ref{fig:logg_teff}, with confidence intervals calculated from the probability $p\propto\exp(-\chi^2/2)$. As the $\chi^2$ does not incorporate any model uncertainties, these confidence intervals only represent the uncertainty on these parameters for this specific model.

\subsection{Comparison with combined evolutionary and atmospheric models}
\begin{figure*}
\epsscale{1.17}
\plotone{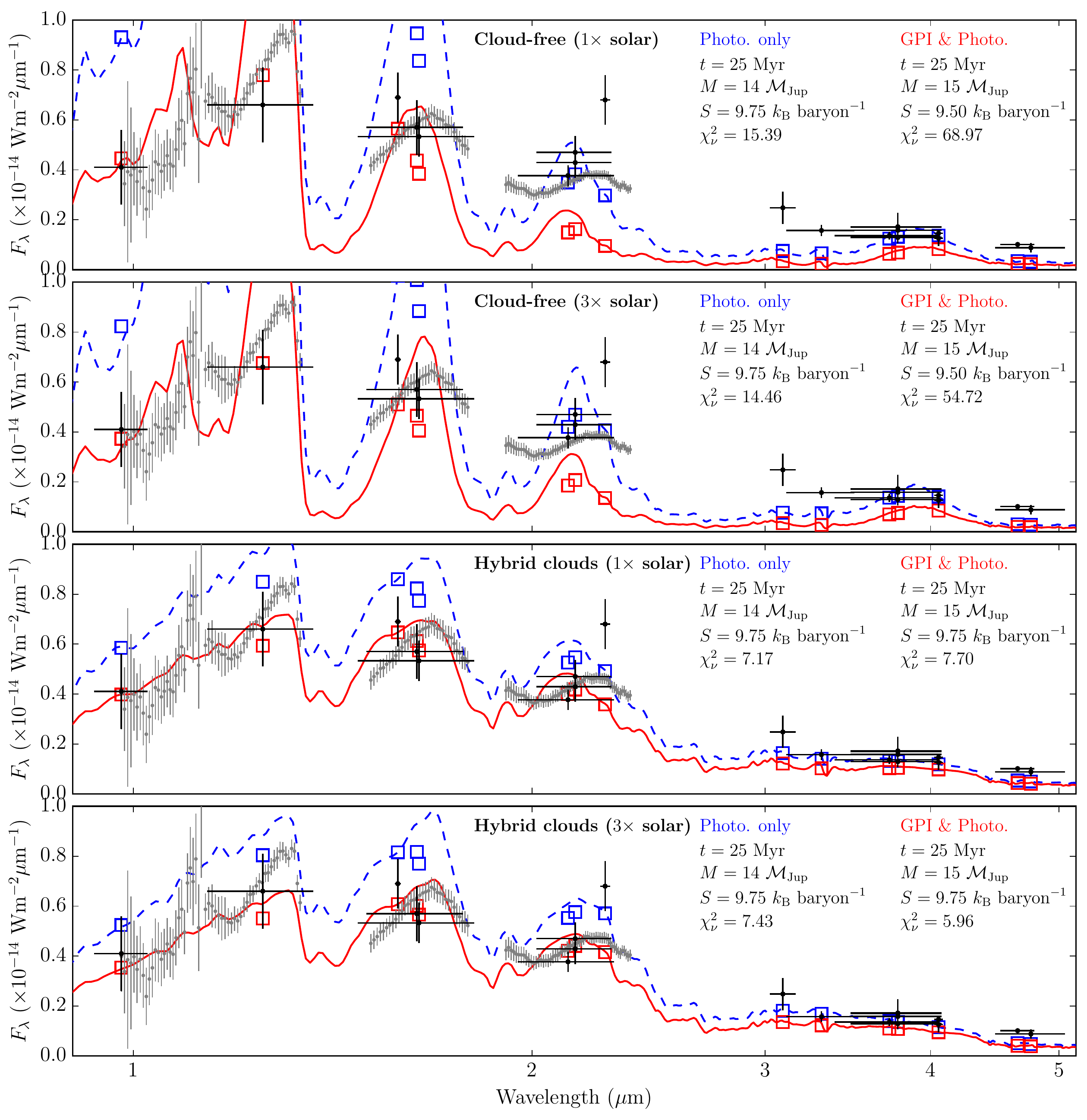}
\caption{The best fit model within each of the four \citet{SpiegelBurrows2012} grids found using only the photometric measurements (blue dashed curve) and using both the photometric and spectroscopic measurements (red solid curve) of $\beta$~Pic~b. The photometric measurements of $\beta$~Pic~b compiled by \citet{Morzinski2015} are plotted as black points, while the GPI spectra presented in this study are plotted as light gray points. Synthetic photometry (open blue and red squares) was computed for each model using the filter profiles shown in Figure~\ref{fig:SED}.}
\label{fig:SED_entropy}
\end{figure*}
\begin{figure*}
\epsscale{1.17}
\plotone{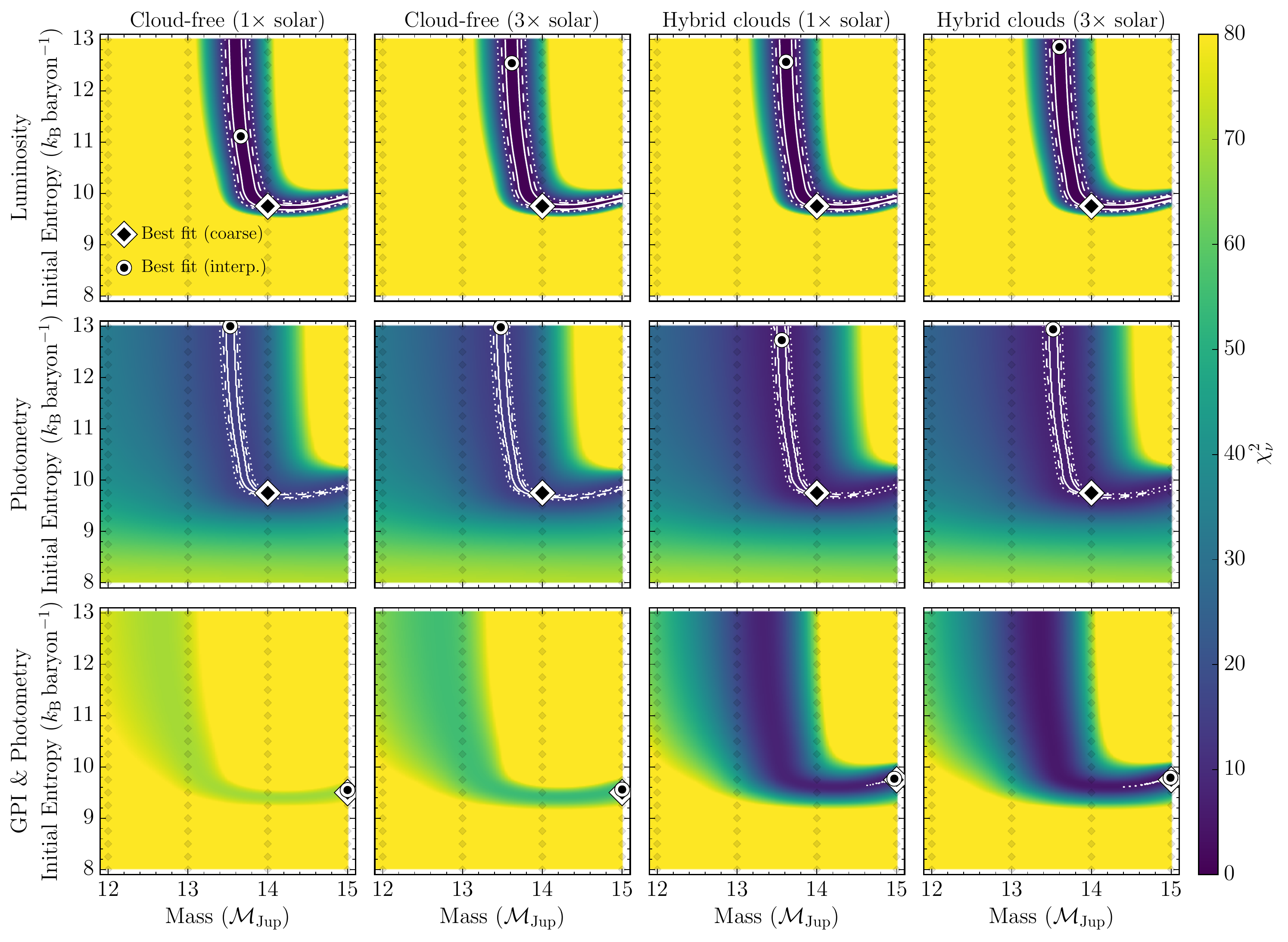}
\caption{Goodness of fit statistic ($\chi^2_{\nu}$) for the cloud-free and hybrid clouds models with both solar and super solar metallicity \citep{SpiegelBurrows2012} fit to the luminosity (top row), photometry (middle row), and photometry and spectroscopy (bottom row) of $\beta$~Pic~b. Model fluxes were interpolated between the grid points (small grey diamonds) to a resolution of 0.01\,\mjup\ in mass and 0.0025\,$k_{\rm B}$/baryon in initial entropy. The best fit model within the original grid is indicated by a large diamond, with the best fit model within the interpolated grid indicated by a circle. The solid, dashed, and dotted contours indicates the 1, 2, and 3\,$\sigma$ confidence interval derived from the $\chi^2$ surface. The poor quality of the fit of these models to the GPI spectrum (bottom row) leads to extremely small confidence intervals as the $\chi^2$ does not incorporate any model uncertainty.}
\label{fig:entropy_mass}
\end{figure*}
The observed SED was also compared with the combined evolutionary and atmospheric models of \citet{SpiegelBurrows2012}\footnote{{\tt http://www.astro.princeton.edu/\~{}burrows/}} using the same fitting procedure as with the previous grids. These models differ from the atmosphere-only models in that the grid was computed in terms of the mass and initial entropy of the planet, rather than the effective temperature and surface gravity. The \citet{SpiegelBurrows2012} grid is bound by the canonical low-entropy ``cold-start'' (8\,$k_{\rm B}$/baryon), and high-entropy ``hot-start'' (13\,$k_{\rm B}$/baryon) models (c.f. \citealt{Marley:2007bf}), where $k_{\rm B}$ is Boltzmann's constant. Here, the initial entropy describes how efficiently heat was radiated away from the forming planet during accretion; formation through gravitational instability may result in a significantly higher initial entropy than formation through core accretion. These evolutionary models were then coupled with an atmospheric model \citep{Hubeny:2003eb,Burrows:2006ia} to create synthetic spectra at each grid point. The atmospheric model used either a solar ($1\times$) or super-solar ($3\times$) metallicity, and either cloud-free or with a linear superposition of cloudy and cloud-free models (hybrid clouds). In total, four grids of synthetic spectra were compared to the SED of $\beta$~Pic~b, spanning this range of atmospheric properties. As the age of $\beta$~Pic~b is well-constrained, the SED of the planet was only fit to the 25\,Myr models within each of the four grids.

As with the fits to the atmospheric models in Section~\ref{sec:atmo_models}, two fits of the SED of $\beta$~Pic~b were made to each grid. The first using only the photometry presented in \citet{Morzinski2015}, and the second combining this with the GPI spectrum presented in this study. The results from these fits are given in Table~\ref{tbl:best_fit_results_entropy} and the best fit spectrum for each of the different models for both photometry only and GPI spectrum and photometry are shown in Figure~\ref{fig:SED_entropy}. The results of the fit to only the photometry of $\beta$~Pic~b are consistent with that of \citet{Morzinski2015}, with a best fit model at a mass of 14.0\,\mjup\ and an initial entropy of 9.75\,$k_{\rm B}$/baryon for each grid. Including the GPI spectrum slightly changed the best fit in each case, to a lower initial entropy for the cloud-free models, and to a higher mass for both the cloud-free and hybrid cloud models. In general, the quality of the fit to the spectrum was poor, with a minimum $\chi^2_{\nu}$ of 6.0 (Figure~\ref{fig:SED_entropy}), compared with a minimum $\chi^2_{\nu}$ of 1.8 for the model atmosphere fits presented in Section~\ref{sec:atmo_models}. Including the GPI spectrum in the fit significantly increases the $\chi^2_{\nu}$ for the cloud-free models, and as such they are not discussed further.

This process was repeated on an interpolated version of the grid to explore the effects of the finitely sampled grid on the results. As in Section~\ref{sec:atmo_models}, a $\chi^2_{\nu}$ surface was calculated for each model grid for both of the data sets. These surfaces, shown in Figure~\ref{fig:entropy_mass}, suggest that the global minimum may have been missed by \citet{Morzinski2015} due to the spacing of the grid points in mass. Using only the photometry, we find a minimum at a significantly higher initial entropy of 13\,$k_{\rm B}$/baryon and a lower mass of 13.5\,\mjup, compared with 9.75\,$k_{\rm B}$/baryon and 14.0\,\mjup\ reported by \citet{Morzinski2015}. The 1\,$\sigma$ confidence interval extends between 10--13\,$k_{\rm B}$/baryon, but is tightly constrained in terms of mass. The fits to the hybrid cloud models including the GPI spectrum are consistent with those from the coarse grid described previously, however the $\chi^2_{\nu}$ surface is similar to that from the photometry-only fit. While the minimum is at an intermediate entropy (9.75\,$k_{\rm B}$/baryon) and high mass (15\,\mjup), this extends to lower masses (13.5\,\mjup) at a range of entropies (10--13\,$k_{\rm B}$/baryon). This complex minimum is also seen when fitting the empirical luminosity of $\beta$~Pic~b given in Section~\ref{sec:bol_lum} to the luminosity of each model grid calculated by integrating the synthetic spectra (Figure~\ref{fig:entropy_mass}, top row). These higher initial entropies are consistent with previous comparisons to evolutionary models showing that the initial entropy is higher than 10.5\,$k_{\rm B}$/baryon at the 95\,\% confidence level \citep{Bonnefoy2014}. 

\section{Conclusion}\label{sec:Discussion}
We present the spectrum of $\beta$~Pic~b in {\it Y}, {\it J}, {\it H}, and {\it K} bands as observed with the Gemini Planet Imager between 2013 and 2016 using images which were taken as part of the verification and commissioning process, as part of an astrometric monitoring program, and as part of a Gemini Large and Long Program using GPI. Not all of the presented data was originally intended to be used for spectral extraction of the planet, but it is of sufficient quality and is valuable as it improves our understanding of the emission spectrum of $\beta$~Pic~b. Using the standard GPI data reduction pipeline and KLIP-FM to extract the spectrum, we recover the planet at a high SNR in $Y$, $J$, $H$, and $K$ bands allowing a nearly full sample across the near-IR.

We compare the spectral energy distribution of $\beta$~Pic~b to that of young, cool, low-surface gravity brown dwarfs, and to several grids of model atmospheres that are valid over the temperature and surface gravity range expected for these objects. Compared with the near-infrared spectra of brown dwarfs in young moving groups and the field, we find that the best fit spectra are those of young low-gravity objects. Of all the objects compared the spectrum of $\beta$~Pic~b best matches that of 2MASS~J03552337+1133437, a confirmed member of the $149^{+51}_{-19}$\,Myr AB Doradus moving group that exhibits strong indicators of low surface gravity \citep{Faherty:2013bc,Liu:2013ej,Gagne:2015dc}. Based on our fits to the low-gravity standards of \citet{Allers2013}, we adopt a spectral type and gravity classification of L$2\gamma\pm1$ for $\beta$~Pic~b.

Combining the GPI spectrum with literature photometry spanning from $Y_{S}$ (0.985\,\micron) to $M^{\prime}$ (4.72\,\micron), we directly measure the bolometric luminosity of the planet to be $\log L_{\rm bol}/{\mathcal L}_{\odot}=-3.76\pm0.02$\,[dex], consistent with previous estimates derived from photometry alone \citep{Morzinski2015}. Comparing to ``hot-start'' evolutionary models \citet{2003A&A...402..701B} yields model-dependent estimates for the physical properties of $\beta$~Pic~b of $M=12.9\pm0.2$\,\mjup, $T_{\rm eff} = 1724\pm15$\,K, $R = 1.46\pm0.01$\,\rjup, and $\log g = 4.18\pm0.01$\,[dex]. The full SED of $\beta$~Pic~b was also compared to atmospheric and evolutionary model grids spanning a range of atmospheric properties and formation scenarios. The best atmospheric fits we find are to a {\sc Drift Phoenix} model atmosphere with an \teff$ = 1700$\,K, \logg$=4.0$\,[dex], and $R=1.41$\,\rjup ($\chi^{2}_{\nu} = 1.81$). These values are consistent with those derived from the bolometric luminosity, and with empirical spectral type to effective temperature relations derived for young low-gravity brown dwarfs (e.g., \citealp{Faherty:2016fx}).

Comparing to the combined atmospheric and evolutionary models of \citet{SpiegelBurrows2012} yielded a best fit at a mass of 15\,\mjup and an intermediate entropy of 9.75 $k_{\rm B}$/baryon, with models including a cloudy atmosphere being strongly preferred over those with a clear atmosphere. While the best fit was found at an initial entropy that is intermediate to the predictions of the ``cold-start'' and ``hot-start'' formation scenarios, the $\chi^2_{\nu}$ surface for the interpolated version of the grid has a complex structure with a minimum extending to lower masses ($\sim13.5$\,\mjup) at a range of initial entropy values between between $\sim$10 and 13 $k_{\rm B}$/baryon, the higher value being similar to that predicted by the ``hot-start`` formation scenarios. Although the points on the finer grid are based on an interpolation of the coarse grid, this analysis suggests that the choice of grid point location and spacing may significantly impact the resulting best fit. If the grid were sampled more finely or shifted by 0.5\,\mjup, and assuming the interpolated points are a reasonable representation of the ``true'' model with those parameters, \citet{Morzinski2015} would have reported a higher entropy as the best fit model.

The empirical bolometric luminosity presented here combined with the dynamical mass constraints from \citet{Lagrange2012}, and the comparison to the atmospheric and evolutionary models of \citet{SpiegelBurrows2012} both suggest a ``hot-start'' high-entropy formation scenario for $\beta$~Pic~b, and are consistent with the prediction that ``cold-start'' low-entropy formation is an unlikely formation mechanism for wide-orbit giant planets \citep[e.g.,][]{Marleau2017}. As $\beta$~Pic~b heads towards maximum elongation in 2023 it will become separated enough from its host star to be resolved by the near- and mid-IR instruments on the upcoming {\it James Webb Space Telescope}. Combining the measurements presented here with mid-IR spectroscopy would provide further insight into the atmospheric properties and evolutionary history of the planet. Interpretation of a well-sampled SED spanning over a decade in wavelength would be extremely well-suited for retrieval techniques \citep[e.g.,][]{Burningham2017} rather than by fitting to finitely-spaced model grids.

\acknowledgments

The Gemini Planet Imager collaboration would like to acknowledge the memory of Leslie Saddlemyer of Canada’s National Research Council, who passed away in January of 2017. Les served as the systems engineer and opto-mechanical lead for GPI during its design, construction, and he deserves great credit for the capabilities of our instrument and team.

We thank the anonymous referee for the helpful comments that improved the quality of this work. The authors would like to acknowledge the financial support of the Gemini Observatory, the Dunlap Institute, University of Toronto, the NSF Center for Adaptive Optics at UC Santa Cruz, the NSF (AST-0909188; AST-1211562, AST-1405505), NASA Origins (NNX11AD21G; NNX10AH31G, NNX14AC21G, NNX15AC89G), and NASA NExSS (NNX15AD95G), the University of California Office of the President (LFRP-118057), and the Science and Technology Facilities Council (ST/H002707/1). Portions of this work were performed under the auspices of the U.S. Department of Energy by Lawrence Livermore National Laboratory under Contract DE-AC52-07NA27344 and under contract with the California Institute of Technology/Jet Propulsion Laboratory funded by NASA through the Sagan Fellowship Program executed by the NASA Exoplanet Science Institute. This work is supported by the NASA Exoplanets Research Program (XRP) by cooperative agreement NNX16AD44G. Support for this work was provided by NASA through Hubble Fellowship grant 51378.01-A awarded by the Space Telescope Science Institute, which is operated by the Association of Universities for Research in Astronomy, Inc., for NASA, under contract NAS5-26555. This work benefited from NASA’s Nexus for Exoplanet System Science (NExSS) research coordination network sponsored by NASA’s Science Mission Directorate. This research has benefited from the SpeX Prism Library (and/or SpeX Prism Library Analysis Toolkit), maintained by Adam Burgasser at http://www.browndwarfs.org/spexprism, the IRTF Spectral Library, maintained by Michael Cushing, and the Montreal Brown Dwarf and Exoplanet Spectral Library, maintained by Jonathan Gagn{\'e}.

\facility{Gemini South (GPI)}.


\clearpage
~

\clearpage

\begin{deluxetable*}{llcrr}
\tablecaption{GPI observations of $\beta$~Pic\label{tbl:observations}} 
\tablewidth{0pt} 
\tablehead{ 
\colhead{Date} & \colhead{\shortstack{Observing \\ Mode}} & \colhead{\shortstack{Exposure \\ Time (s)}} & \colhead{\shortstack{Parallactic \\ Rotation (\degr)}}  & \colhead{\shortstack{DIMM \\ Seeing (\arcsec)}}}
 \startdata 
2013-11-16\tablenotemark{d,f} & K1-Spec. & 1789 & 26 & 1.09 \\
2013-11-16\tablenotemark{d,f} & K2-Spec. & 1253 & 18 & 0.93 \\
2013-11-18\tablenotemark{a,c,d,e,f} & H-Spec.  & 2446 & 32 & 0.68  \\
2013-12-10\tablenotemark{d,f} & H-Spec.  & 1312 & 38 & 0.77 \\
2013-12-10\tablenotemark{b,d,e,f} & J-Spec.  & 1597 & 18 & 0.70 \\
2013-12-11\tablenotemark{d,f} & H-Spec.  & 556  & 64 & 0.46 \\
2014-03-23\tablenotemark{d,f} & K1-Spec. & 1133 & 47 & 0.47 \\ 
2014-03-26\tablenotemark{f} & K2-Spec. & 2923 & 26 & 0.86\\
2014-11-08\tablenotemark{d} & H-Spec.  & 2147 & 25 & 0.77 \\ 
2015-12-05\tablenotemark{h} & Y-Spec.  & 2002 & 37 & 1.12 \\
\enddata 
\tablenotetext{a}{This data set was originally astrometrically published by \citet{Macintosh2014}} 
\tablenotetext{b}{This data set was origninally published by \citet{Bonnefoy2014}} 
\tablenotetext{c}{This data set for a spectrum was published by \citet{Chilcote2015}}
\tablenotetext{d}{This data set was originally astrometrically published by \citet{MillarBlanchaer2015}}
\tablenotetext{e}{CCR power state was changed during observations}
\tablenotetext{f}{Observations taken during GPI verification \& commitioning tests}
\tablenotetext{g}{AO performance parameters adjusted during GPI verification \& commitioning tests}
\tablenotetext{h}{Data part of Gemini's Large and Long program (GS-2015B-LP-6)}
\end{deluxetable*}

\clearpage

\begin{table*}
	\caption{Measured $\beta$~Pic~b Parameters}
	\nopagebreak
	\centering
	\begin{tabular}{llllll}
		\hline
		\hline
		& \teff	& \logg	& Radius  & Mass & Init.\ Spec.\ Entropy \\
		Reference & (K) & (cgs) & (\rjup) & (\mjup) & ($k_{\rm B}$ baryon$^{-1}$) \\
		\hline
		\citet{Currie2013} & 1575$^*$ & 3.8$\pm$0.2 & 1.65$\pm$0.06 & 6.9$^*$ & $\cdots$ \\
		\citet{Bonnefoy2013} & 1700$\pm$100	& 4.0$\pm$0.5 & 1.5--1.6$^*$  & 9--10 & $\geq$9.3 \\
		\citet{Bonnefoy2014} & 1650$\pm$150	& $<$4.7 & 1.5$\pm$0.2 & $<$20	& $>$10.5 \\
		\citet{Chilcote2015} & 1600--1700 & 3.5--4.5 & $\cdots$ & $\cdots$ & $\cdots$ \\
		\citet{Baudino2015} & 1550$\pm$150 & 3.5$\pm$1 & 1.76$\pm$0.24 & 4.0$^*$ & $\cdots$ \\
		\citet{Morzinski2015} & 1708$\pm$23 & 4.2 & 1.45$\pm$0.02 & 12.7$\pm$0.3 & 9.75 \\
		This work (Bolometric) & 1724$\pm$15 & 4.18$\pm$0.01 & 1.46$\pm$0.01 & 12.9$\pm$0.2  & \\
		This work (Spectrophotometry)$^\dagger$ & 1700 & 4.0 & 1.41 & 15.0 & 9.75\\
		\hline
		\multicolumn{6}{l}{}	\\     
		\multicolumn{6}{l}{$^*$ Value calculated in \citet{Morzinski2015}.} \\
		\multicolumn{6}{l}{$^\dagger$ Best fit from Drift-Phoenix and \citet{SpiegelBurrows2012} models}\\
	\end{tabular}
	\label{tbl:allworks}
\end{table*}

\clearpage

\begin{deluxetable*}{llcccc|cccc}
\tabletypesize{\scriptsize} 
\tablecaption{Best-fit atmospheric models} 
\tablewidth{0pt} 
\tablehead{
\colhead{Grid Name} & \colhead{Data Used} & \multicolumn{4}{c}{Grid Points} & \multicolumn{4}{c}{Interpolated Grid} \\
 & & \colhead{\teff} & \colhead{\logg} & \colhead{Radius} & \colhead{$\chi^{2}_{\nu}$} & \colhead{\teff} & \colhead{\logg} & \colhead{Radius} & \colhead{$\chi^{2}_{\nu}$} \\
 & & \colhead{(K)} & \colhead{([dex])} & \colhead{(\rjup)} & & \colhead{(K)} & \colhead{([dex])} & \colhead{(\rjup)}
}
\startdata
{\sc Ames-Dusty} & Photometry Only & 1700 & 3.5 & 1.31 & 2.66 & 1704 & 3.50 & 1.31 & 2.66 \\
 & GPI Spectrum \& Photometry & 1800 & 3.5 & 1.17 & 3.49 & 1706 & 4.50 & 1.18 & 3.45 \\
{\sc BT-Settl} (2015) & Photometry Only & 1800 & 3.0 & 1.38 & 2.99 & 1781 & 3.26 & 1.34 & 2.63 \\
 & GPI Spectrum \& Photometry & 1800 & 3.5 & 1.22 & 3.17 & 1815 & 3.29 & 1.25 & 3.04 \\
{\sc Drift Phoenix} & Photometry Only & 1700 & 3.5 & 1.41 & 1.55 & 1708 & 3.66 & 1.41 & 1.54 \\
 & GPI Spectrum \& Photometry & 1700 & 4.0 & 1.41 & 1.81 & 1651 & 3.00 & 1.58 & 1.21 \\
\enddata
\label{tbl:best_fit_results}
\end{deluxetable*}

\clearpage

\begin{deluxetable*}{llccc|ccc}
\tabletypesize{\scriptsize} 
\tablecaption{Best-fit combined evolutionary and atmospheric models} 
\tablewidth{0pt} 
\tablehead{
\colhead{Grid Name} & \colhead{Data Used} & \multicolumn{3}{c}{Grid Points} & \multicolumn{3}{c}{Interpolated Grid} \\
 & & \colhead{Mass} & \colhead{Initial Entropy} & \colhead{$\chi^{2}_{\nu}$}  & \colhead{Mass} & \colhead{Initial Entropy} &  \colhead{$\chi^{2}_{\nu}$} \\
 & & \colhead{(\mjup)} & \colhead{($k_{\rm B}$/baryon)}& & \colhead{(\mjup)} & \colhead{($k_{\rm B}$/baryon)}
}
\startdata
Cloud-free ($1\times$ solar) & Photometry Only & 14.0  & 9.75 & 15.39 & 13.53 & 13.00 & 15.18\\
 &  Spectrum \& Photometry & 15.0 & 9.50 & 68.97 & 15.00  & 9.55  & 68.45 \\
Cloud-free ($3\times$ solar) & Photometry Only & 14.0 & 9.75 & 14.46 & 13.48 & 12.98 & 13.91\\
 &  Spectrum \& Photometry & 15.0 & 9.50 & 54.72 & 15.00 & 9.56  & 53.52 \\
Hybrid clouds ($1\times$ solar) & Photometry Only & 14.0  & 9.75 & 7.17 & 13.56 & 12.73  & 7.00 \\
 &  Spectrum \& Photometry & 15.0 & 9.75  & 7.70  & 14.97 & 9.77 & 7.02 \\
Hybrid clouds ($3\times$ solar) & Photometry Only & 14.0  & 9.75  & 7.43 & 13.52 & 12.94 & 7.17 \\
 &  Spectrum \& Photometry & 15.0 & 9.75 & 5.96 & 14.99 & 9.79 & 4.87 \\
\enddata
\label{tbl:best_fit_results_entropy}
\end{deluxetable*}

\begin{thebibliography}{27}
\expandafter\ifx\csname natexlab\endcsname\relax\def\natexlab#1{#1}\fi

\bibitem[{{Amari} {et~al.}(1992){Amari}, {Hoppe}, {Zinner}, \&
  {Lewis}}]{1992ApJ...394L..43A}
{Amari}, S., {Hoppe}, P., {Zinner}, E., \& {Lewis}, R.~S. 1992, \apjl, 394, L43

\bibitem[{{Amari} \& {Zinner}(1997)}]{1997ails.conf..287A}
{Amari}, S. \& {Zinner}, E. 1997, in Astrophysical Implications of the
  Laboratory Study of Presolar Materials, 287--305

\bibitem[{{Amari} {et~al.}(1996){Amari}, {Zinner}, \&
  {Lewis}}]{1996ApJ...470L.101A}
{Amari}, S., {Zinner}, E., \& {Lewis}, R.~S. 1996, \apjl, 470, L101

\bibitem[{{Clayton}(1975)}]{1975Natur.257...36C}
{Clayton}, D.~D. 1975, \nat, 257, 36

\bibitem[{{Clayton}(1989)}]{1989ApJ...340..613C}
---. 1989, \apj, 340, 613

\bibitem[{{Clayton} {et~al.}(1996){Clayton}, {Amari}, \&
  {Zinner}}]{1996ApSpSci...251..355C}
{Clayton}, D.~D., {Amari}, S., \& {Zinner}, E. 1996, \apss, 251, 355

\bibitem[{{Clayton} {et~al.}(1997){Clayton}, {Arnett}, {Kane}, \&
  {Meyer}}]{1997ApJ...486..824C}
{Clayton}, D.~D., {Arnett}, D., {Kane}, J., \& {Meyer}, B.~S. 1997, \apj, 486,
  824

\bibitem[{{Clayton} {et~al.}(1999){Clayton}, {Liu}, \&
  {Dalgarno}}]{1999Sci...283.1290C}
{Clayton}, D.~D., {Liu}, W., \& {Dalgarno}, A. 1999, Science, 283, 1290

\bibitem[{{Clayton} {et~al.}(1995){Clayton}, {Meyer}, {Sanderson}, {Russell},
  \& {Pillinger}}]{1995ApJ...447..894C}
{Clayton}, D.~D., {Meyer}, B.~S., {Sanderson}, C.~I., {Russell}, S.~S., \&
  {Pillinger}, C.~T. 1995, \apj, 447, 894

\bibitem[{{Heymann}(1983)}]{1983ApJ...267..747H}
{Heymann}, D. 1983, \apj, 267, 747

\bibitem[{{Heymann} \& {Dziczkaniec}(1979)}]{1979LPSC...10.1943H}
{Heymann}, D. \& {Dziczkaniec}, M. 1979, in Lunar and Planetary Science
  Conference, 10th, Houston, Tex., March 19-23, 1979, Proceedings. Volume 2.
  (A80-23617 08-91) New York, Pergamon Press, Inc., 1979, p. 1943-1959.,
  Vol.~10, 1943--1959

\bibitem[{{Heymann} \& {Dziczkaniec}(1980)}]{1980LPSC...11.1179H}
{Heymann}, D. \& {Dziczkaniec}, M. 1980, in Lunar and Planetary Science
  Conference, 11th, Houston, TX, March 17-21, 1980, Proceedings. Volume 2.
  (A82-22296 09-91) New York, Pergamon Press, 1980, p. 1179-1213., Vol.~11,
  1179--1213

\bibitem[{{Hoppe} {et~al.}(1994){Hoppe}, {Amari}, {Zinner}, {Ireland}, \&
  {Lewis}}]{1994ApJ...430..870H}
{Hoppe}, P., {Amari}, S., {Zinner}, E., {Ireland}, T., \& {Lewis}, R.~S. 1994,
  \apj, 430, 870

\bibitem[{{Hoppe} {et~al.}(1995){Hoppe}, {Amari}, {Zinner}, \&
  {Lewis}}]{1995GeCoA..59.4029H}
{Hoppe}, P., {Amari}, S., {Zinner}, E., \& {Lewis}, R.~S. 1995, \gca, 59, 4029

\bibitem[{{Hoppe} {et~al.}(1993){Hoppe}, {Geiss}, {Buehler}, {Neuenschwander},
  {Amari}, \& {Lewis}}]{1993GeCoA..57.4059H}
{Hoppe}, P., {Geiss}, J., {Buehler}, F., {Neuenschwander}, J., {Amari}, S., \&
  {Lewis}, R.~S. 1993, \gca, 57, 4059

\bibitem[{{Hoppe} {et~al.}(1996{\natexlab{a}}){Hoppe}, {Strebel}, {Eberhardt},
  {Amari}, \& {Lewis}}]{1996Sci...272.1314H}
{Hoppe}, P., {Strebel}, R., {Eberhardt}, P., {Amari}, S., \& {Lewis}, R.~S.
  1996{\natexlab{a}}, Science, 272, 1314

\bibitem[{{Hoppe} {et~al.}(1996{\natexlab{b}}){Hoppe}, {Strebel}, {Eberhardt},
  {Amari}, \& {Lewis}}]{1996GeCoA..60..883H}
---. 1996{\natexlab{b}}, \gca, 60, 883

\bibitem[{{Howard} {et~al.}(1992){Howard}, {Meyer}, \&
  {Clayton}}]{1992Metic..27..404H}
{Howard}, W.~M., {Meyer}, B.~S., \& {Clayton}, D.~D. 1992, Meteoritics, 27, 404

\bibitem[{{Meyer}(1995)}]{1995ApJ...449L..55M}
{Meyer}, B.~S. 1995, \apjl, 449, L55

\bibitem[{{Nicolussi} {et~al.}(1998){Nicolussi}, {Pellin}, {Lewis}, {Davis},
  {Clayton}, \& {Amari}}]{1998ApJ...504..492N}
{Nicolussi}, G.~K., {Pellin}, M.~J., {Lewis}, R.~S., {Davis}, A.~M., {Clayton},
  R.~N., \& {Amari}, S. 1998, \apj, 504, 492

\bibitem[{{Nittler} {et~al.}(1996){Nittler}, {Amari}, {Zinner}, {Woosley}, \&
  {Lewis}}]{1996ApJ...462L..31N}
{Nittler}, L.~R., {Amari}, S., {Zinner}, E., {Woosley}, S.~E., \& {Lewis},
  R.~S. 1996, \apjl, 462, L31

\bibitem[{{Nittler} {et~al.}(1995){Nittler}, {Hoppe}, {Alexander}, {Amari},
  {Eberhardt}, {Gao}, {Lewis}, {Strebel}, {Walker}, \&
  {Zinner}}]{1995ApJ...453L..25N}
{Nittler}, L.~R., {Hoppe}, P., {Alexander}, C. M.~O., {Amari}, S., {Eberhardt},
  P., {Gao}, X., {Lewis}, R.~S., {Strebel}, R., {Walker}, R.~M., \& {Zinner},
  E. 1995, \apjl, 453, L25

\bibitem[{{Pellin} {et~al.}(2000){Pellin}, {Calaway}, {Davis}, {Lewis},
  {Amari}, \& {Clayton}}]{2000LPI....31.1917P}
{Pellin}, M.~J., {Calaway}, W.~F., {Davis}, A.~M., {Lewis}, R.~S., {Amari}, S.,
  \& {Clayton}, R.~N. 2000, in Lunar and Planetary Science Conference, Vol.~31,
  1917

\bibitem[{{Pellin} {et~al.}(1999){Pellin}, {Davis}, {Lewis}, {Amari}, \&
  {Clayton}}]{1999LPI....30.1969P}
{Pellin}, M.~J., {Davis}, A.~M., {Lewis}, R.~S., {Amari}, S., \& {Clayton},
  R.~N. 1999, in Lunar and Planetary Science Conference, Vol.~30, 1969

\bibitem[{{Srinivasan} \& {Anders}(1978)}]{1978Sci...201...51S}
{Srinivasan}, B. \& {Anders}, E. 1978, Science, 201, 51

\bibitem[{{Travaglio} {et~al.}(1999){Travaglio}, {Gallino}, {Amari}, {Zinner},
  {Woosley}, \& {Lewis}}]{1999ApJ...510..325T}
{Travaglio}, C., {Gallino}, R., {Amari}, S., {Zinner}, E., {Woosley}, S., \&
  {Lewis}, R.~S. 1999, \apj, 510, 325

\bibitem[{{Zinner}(1997)}]{1997ails.conf....3Z}
{Zinner}, E. 1997, in Astrophysical Implications of the Laboratory Study of
  Presolar Materials, 3--26

\end{thebibliography}


\begin{thebibliography}{}
\expandafter\ifx\csname natexlab\endcsname\relax\def\natexlab#1{#1}\fi

\bibitem[{{Absil} {et~al.}(2013){Absil}, {Milli}, {Mawet}, {Lagrange},
  {Girard}, {Chauvin}, {Boccaletti}, {Delacroix}, \& {Surdej}}]{Absil2013}
{Absil}, O., {Milli}, J., {Mawet}, D., {et~al.} 2013, \aap, 559, L12

\bibitem[{{Allard} {et~al.}(2001){Allard}, {Hauschildt}, {Alexander},
  {Tamanai}, \& {Schweitzer}}]{Allard2001}
{Allard}, F., {Hauschildt}, P.~H., {Alexander}, D.~R., {Tamanai}, A., \&
  {Schweitzer}, A. 2001, \apj, 556, 357

\bibitem[{{Allard} {et~al.}(2012){Allard}, {Homeier}, \&
  {Freytag}}]{Allard2012}
{Allard}, F., {Homeier}, D., \& {Freytag}, B. 2012, Royal Society of London
  Philosophical Transactions Series A, 370, 2765

\bibitem[{{Allers} \& {Liu}(2013)}]{Allers2013}
{Allers}, K.~N., \& {Liu}, M.~C. 2013, Astrophysical Journal, 772, 79

\bibitem[{{Augereau} {et~al.}(2001){Augereau}, {Nelson}, {Lagrange},
  {Papaloizou}, \& {Mouillet}}]{Augereau2001}
{Augereau}, J.~C., {Nelson}, R.~P., {Lagrange}, A.~M., {Papaloizou}, J.~C.~B.,
  \& {Mouillet}, D. 2001, Astronomy and Astrophysics, 370, 447

\bibitem[{{Baraffe} {et~al.}(2003){Baraffe}, {Chabrier}, {Barman}, {Allard}, \&
  {Hauschildt}}]{2003A&A...402..701B}
{Baraffe}, I., {Chabrier}, G., {Barman}, T.~S., {Allard}, F., \& {Hauschildt},
  P.~H. 2003, \aap, 402, 701

\bibitem[{{Baudino} {et~al.}(2015){Baudino}, {B{\'e}zard}, {Boccaletti},
  {Bonnefoy}, {Lagrange}, \& {Galicher}}]{Baudino2015}
{Baudino}, J.-L., {B{\'e}zard}, B., {Boccaletti}, A., {et~al.} 2015, \aap, 582,
  A83

\bibitem[{{Bell} {et~al.}(2015){Bell}, {Mamajek}, \& {Naylor}}]{Bell2015}
{Bell}, C.~P.~M., {Mamajek}, E.~E., \& {Naylor}, T. 2015, \mnras, 454, 593

\bibitem[{{Boccaletti} {et~al.}(2013){Boccaletti}, {Lagrange}, {Bonnefoy},
  {Galicher}, \& {Chauvin}}]{Boccaletti2013}
{Boccaletti}, A., {Lagrange}, A.-M., {Bonnefoy}, M., {Galicher}, R., \&
  {Chauvin}, G. 2013, Astronomy and Astrophysics, 551, L14

\bibitem[{{Bonnefoy} {et~al.}(2011){Bonnefoy}, {Lagrange}, {Boccaletti},
  {Chauvin}, {Apai}, {Allard}, {Ehrenreich}, {Girard}, {Mouillet}, {Rouan},
  {Gratadour}, \& {Kasper}}]{Bonnefoy2011}
{Bonnefoy}, M., {Lagrange}, A.-M., {Boccaletti}, A., {et~al.} 2011, \aap, 528,
  L15

\bibitem[{{Bonnefoy} {et~al.}(2013){Bonnefoy}, {Boccaletti}, {Lagrange},
  {Allard}, {Mordasini}, {Beust}, {Chauvin}, {Girard}, {Homeier}, {Apai},
  {Lacour}, \& {Rouan}}]{Bonnefoy2013}
{Bonnefoy}, M., {Boccaletti}, A., {Lagrange}, A.-M., {et~al.} 2013, Astronomy
  and Astrophysics, 555, A107

\bibitem[{{Bonnefoy} {et~al.}(2014){Bonnefoy}, {Marleau}, {Galicher}, {Beust},
  {Lagrange}, {Baudino}, {Chauvin}, {Borgniet}, {Meunier}, {Rameau},
  {Boccaletti}, {Cumming}, {Helling}, {Homeier}, {Allard}, \&
  {Delorme}}]{Bonnefoy2014}
{Bonnefoy}, M., {Marleau}, G.-D., {Galicher}, R., {et~al.} 2014, \aap, 567, L9

\bibitem[{{Burgasser}(2014)}]{2014ASInC..11....7B}
{Burgasser}, A.~J. 2014, in Astronomical Society of India Conference Series,
  Vol.~11, Astronomical Society of India Conference Series

\bibitem[{{Burgasser} {et~al.}(2010){Burgasser}, {Cruz}, {Cushing}, {Gelino},
  {Looper}, {Faherty}, {Kirkpatrick}, \& {Reid}}]{2010ApJ...710.1142B}
{Burgasser}, A.~J., {Cruz}, K.~L., {Cushing}, M., {et~al.} 2010, \apj, 710,
  1142

\bibitem[{Burgasser {et~al.}(2006)Burgasser, Geballe, Leggett, Kirkpatrick, \&
  Golimowski}]{Burgasser:2006cf}
Burgasser, A.~J., Geballe, T.~R., Leggett, S.~K., Kirkpatrick, J.~D., \&
  Golimowski, D.~A. 2006, Astrophys. J., 637, 1067

\bibitem[{{Burgasser} {et~al.}(2016){Burgasser}, {Lopez}, {Mamajek},
  {Gagn{\'e}}, {Faherty}, {Tallis}, {Choban}, {Tamiya}, {Escala}, \&
  {Aganze}}]{2016ApJ...820...32B}
{Burgasser}, A.~J., {Lopez}, M.~A., {Mamajek}, E.~E., {et~al.} 2016, \apj, 820,
  32

\bibitem[{{Burningham} {et~al.}(2017){Burningham}, {Marley}, {Line},
  {Visscher}, {Morley}, {Saumon}, {Lupu}, \& {Freedman}}]{Burningham2017}
{Burningham}, B., {Marley}, M.~S., {Line}, M.~R., {et~al.} 2017, ArXiv
  e-prints, arXiv:1701.01257

\bibitem[{{Burrows} {et~al.}(1995){Burrows}, {Saumon}, {Guillot}, {Hubbard}, \&
  {Lunine}}]{Burrows1995}
{Burrows}, A., {Saumon}, D., {Guillot}, T., {Hubbard}, W.~B., \& {Lunine},
  J.~I. 1995, Nature, 375, 299

\bibitem[{Burrows {et~al.}(2006)Burrows, Sudarsky, \& Hubeny}]{Burrows:2006ia}
Burrows, A., Sudarsky, D., \& Hubeny, I. 2006, ApJ, 640, 1063

\bibitem[{{Chabrier} {et~al.}(2000){Chabrier}, {Baraffe}, {Allard}, \&
  {Hauschildt}}]{Chabrier2000}
{Chabrier}, G., {Baraffe}, I., {Allard}, F., \& {Hauschildt}, P. 2000,
  Astrophysical Journal, 542, 464

\bibitem[{{Chauvin} {et~al.}(2005){Chauvin}, {Lagrange}, {Dumas}, {Zuckerman},
  {Mouillet}, {Song}, {Beuzit}, \& {Lowrance}}]{Chauvin2005}
{Chauvin}, G., {Lagrange}, A.-M., {Dumas}, C., {et~al.} 2005, Astronomy and
  Astrophysics, 438, L25

\bibitem[{{Chauvin} {et~al.}(2012){Chauvin}, {Lagrange}, {Beust}, {Bonnefoy},
  {Boccaletti}, {Apai}, {Allard}, {Ehrenreich}, {Girard}, {Mouillet}, \&
  {Rouan}}]{chauvin2012}
{Chauvin}, G., {Lagrange}, A.-M., {Beust}, H., {et~al.} 2012, Astronomy and
  Astrophysics, 542, A41

\bibitem[{{Chilcote} {et~al.}(2015){Chilcote}, {Barman}, {Fitzgerald},
  {Graham}, {Larkin}, {Macintosh}, {Bauman}, {Burrows}, {Cardwell}, {De Rosa},
  {Dillon}, {Doyon}, {Dunn}, {Erikson}, {Gavel}, {Goodsell}, {Hartung},
  {Hibon}, {Ingraham}, {Kalas}, {Konopacky}, {Maire}, {Marchis}, {Marley},
  {Marois}, {Millar-Blanchaer}, {Morzinski}, {Norton}, {Oppenheimer}, {Palmer},
  {Patience}, {Perrin}, {Poyneer}, {Pueyo}, {Rantakyr{\"o}}, {Sadakuni},
  {Saddlemyer}, {Savransky}, {Serio}, {Sivaramakrishnan}, {Song}, {Soummer},
  {Thomas}, {Wallace}, {Wiktorowicz}, \& {Wolff}}]{Chilcote2015}
{Chilcote}, J., {Barman}, T., {Fitzgerald}, M.~P., {et~al.} 2015, \apjl, 798,
  L3

\bibitem[{{Chilcote} {et~al.}(2012){Chilcote}, {Larkin}, {Maire}, {Perrin},
  {Fitzgerald}, {Doyon}, {Thibault}, {Bauman}, {Macintosh}, {Graham}, \&
  {Saddlemyer}}]{Chilcote2012}
{Chilcote}, J.~K., {Larkin}, J.~E., {Maire}, J., {et~al.} 2012, in Society of
  Photo-Optical Instrumentation Engineers (SPIE) Conference Series, Vol. 8446,
  Society of Photo-Optical Instrumentation Engineers (SPIE) Conference Series

\bibitem[{{Crepp} {et~al.}(2015){Crepp}, {Rice}, {Veicht}, {Aguilar}, {Pueyo},
  {Giorla}, {Nilsson}, {Luszcz-Cook}, {Oppenheimer}, {Hinkley}, {Brenner},
  {Vasisht}, {Cady}, {Beichman}, {Hillenbrand}, {Lockhart}, {Matthews},
  {Roberts}, {Sivaramakrishnan}, {Soummer}, \& {Zhai}}]{Crepp2015}
{Crepp}, J.~R., {Rice}, E.~L., {Veicht}, A., {et~al.} 2015, \apjl, 798, L43

\bibitem[{{Cruz} {et~al.}(2009){Cruz}, {Kirkpatrick}, \&
  {Burgasser}}]{2009AJ....137.3345C}
{Cruz}, K.~L., {Kirkpatrick}, J.~D., \& {Burgasser}, A.~J. 2009, \aj, 137, 3345

\bibitem[{{Currie} {et~al.}(2011){Currie}, {Burrows}, {Itoh}, {Matsumura},
  {Fukagawa}, {Apai}, {Madhusudhan}, {Hinz}, {Rodigas}, {Kasper}, {Pyo}, \&
  {Ogino}}]{Currie2011b}
{Currie}, T., {Burrows}, A., {Itoh}, Y., {et~al.} 2011, Astrophysical Journal,
  729, 128

\bibitem[{{Currie} {et~al.}(2013){Currie}, {Burrows}, {Madhusudhan},
  {Fukagawa}, {Girard}, {Dawson}, {Murray-Clay}, {Kenyon}, {Kuchner},
  {Matsumura}, {Jayawardhana}, {Chambers}, \& {Bromley}}]{Currie2013}
{Currie}, T., {Burrows}, A., {Madhusudhan}, N., {et~al.} 2013, Astrophysical
  Journal, 776, 15

\bibitem[{{Cushing} {et~al.}(2005){Cushing}, {Rayner}, \&
  {Vacca}}]{2005ApJ...623.1115C}
{Cushing}, M.~C., {Rayner}, J.~T., \& {Vacca}, W.~D. 2005, \apj, 623, 1115

\bibitem[{{Delorme} {et~al.}(2012){Delorme}, {Gagn{\'e}}, {Malo}, {Reyl{\'e}},
  {Artigau}, {Albert}, {Forveille}, {Delfosse}, {Allard}, \&
  {Homeier}}]{2012A&A...548A..26D}
{Delorme}, P., {Gagn{\'e}}, J., {Malo}, L., {et~al.} 2012, \aap, 548, A26

\bibitem[{Faherty {et~al.}(2013)Faherty, Rice, Cruz, Mamajek, \&
  N{\'u}{\~n}ez}]{Faherty:2013bc}
Faherty, J.~K., Rice, E.~L., Cruz, K.~L., Mamajek, E.~E., \& N{\'u}{\~n}ez, A.
  2013, AJ, 145, 2

\bibitem[{Faherty {et~al.}(2016)Faherty, Riedel, Cruz, Gagn{\'e}, Filippazzo,
  Lambrides, Fica, Weinberger, Thorstensen, Tinney, Baldassare, Lemonier, \&
  Rice}]{Faherty:2016fx}
Faherty, J.~K., Riedel, A.~R., Cruz, K.~L., {et~al.} 2016, ApJS, 225, 10

\bibitem[{{Filippazzo} {et~al.}(2015){Filippazzo}, {Rice}, {Faherty}, {Cruz},
  {Van Gordon}, \& {Looper}}]{2015ApJ...810..158F}
{Filippazzo}, J.~C., {Rice}, E.~L., {Faherty}, J., {et~al.} 2015, \apj, 810,
  158

\bibitem[{Gagn{\'e} {et~al.}(2014)Gagn{\'e}, Lafreni{\`e}re, Doyon, Malo, \&
  Artigau}]{Gagne:2014gp}
Gagn{\'e}, J., Lafreni{\`e}re, D., Doyon, R., Malo, L., \& Artigau, {\'E}.
  2014, ApJ, 783, 121

\bibitem[{Gagn{\'e} {et~al.}(2015)Gagn{\'e}, Faherty, Cruz, Lafreni{\`e}re,
  Doyon, Malo, Burgasser, Naud, Artigau, Bouchard, Gizis, \&
  Albert}]{Gagne:2015dc}
Gagn{\'e}, J., Faherty, J.~K., Cruz, K.~L., {et~al.} 2015, ApJS, 219, 33

\bibitem[{{Golimowski} {et~al.}(2006){Golimowski}, {Ardila}, {Krist},
  {Clampin}, {Ford}, {Illingworth}, {Bartko}, {Ben{\'{\i}}tez}, {Blakeslee},
  {Bouwens}, {Bradley}, {Broadhurst}, {Brown}, {Burrows}, {Cheng}, {Cross},
  {Demarco}, {Feldman}, {Franx}, {Goto}, {Gronwall}, {Hartig}, {Holden},
  {Homeier}, {Infante}, {Jee}, {Kimble}, {Lesser}, {Martel}, {Mei},
  {Menanteau}, {Meurer}, {Miley}, {Motta}, {Postman}, {Rosati}, {Sirianni},
  {Sparks}, {Tran}, {Tsvetanov}, {White}, {Zheng}, \& {Zirm}}]{Golimowski2006}
{Golimowski}, D.~A., {Ardila}, D.~R., {Krist}, J.~E., {et~al.} 2006,
  Astronomical Journal, 131, 3109

\bibitem[{{Graham} {et~al.}(2007){Graham}, {Macintosh}, {Doyon}, {Gavel},
  {Larkin}, {Levine}, {Oppenheimer}, {Palmer}, {Saddlemyer},
  {Sivaramakrishnan}, {Veran}, \& {Wallace}}]{Graham2007}
{Graham}, J.~R., {Macintosh}, B., {Doyon}, R., {et~al.} 2007, ArXiv e-prints,
  arXiv:0704.1454

\bibitem[{{Gray} {et~al.}(2006){Gray}, {Corbally}, {Garrison}, {McFadden},
  {Bubar}, {McGahee}, {O'Donoghue}, \& {Knox}}]{Gray2006}
{Gray}, R.~O., {Corbally}, C.~J., {Garrison}, R.~F., {et~al.} 2006,
  Astronomical Journal, 132, 161

\bibitem[{{Heap} {et~al.}(2000){Heap}, {Lindler}, {Lanz}, {Cornett}, {Hubeny},
  {Maran}, \& {Woodgate}}]{Heap2000}
{Heap}, S.~R., {Lindler}, D.~J., {Lanz}, T.~M., {et~al.} 2000, Astrophysical
  Journal, 539, 435

\bibitem[{{Helling} {et~al.}(2008){Helling}, {Dehn}, {Woitke}, \&
  {Hauschildt}}]{Helling2008}
{Helling}, C., {Dehn}, M., {Woitke}, P., \& {Hauschildt}, P.~H. 2008, \apjl,
  675, L105

\bibitem[{{Helling} \& {Woitke}(2006)}]{Helling2006}
{Helling}, C., \& {Woitke}, P. 2006, \aap, 455, 325

\bibitem[{{Hinkley} {et~al.}(2013){Hinkley}, {Pueyo}, {Faherty}, {Oppenheimer},
  {Mamajek}, {Kraus}, {Rice}, {Ireland}, {David}, {Hillenbrand}, {Vasisht},
  {Cady}, {Brenner}, {Veicht}, {Nilsson}, {Zimmerman}, {Parry}, {Beichman},
  {Dekany}, {Roberts}, {Roberts}, {Baranec}, {Crepp}, {Burruss}, {Wallace},
  {King}, {Zhai}, {Lockhart}, {Shao}, {Soummer}, {Sivaramakrishnan}, \&
  {Wilson}}]{Hinkley2013}
{Hinkley}, S., {Pueyo}, L., {Faherty}, J.~K., {et~al.} 2013, \apj, 779, 153

\bibitem[{Hubeny {et~al.}(2003)Hubeny, Burrows, \& Sudarsky}]{Hubeny:2003eb}
Hubeny, I., Burrows, A., \& Sudarsky, D. 2003, ApJ, 594, 1011

\bibitem[{{Ingraham} {et~al.}(2014){Ingraham}, {Marley}, {Saumon}, {Marois},
  {Macintosh}, {Barman}, {Bauman}, {Burrows}, {Chilcote}, {De Rosa}, {Dillon},
  {Doyon}, {Dunn}, {Erikson}, {Fitzgerald}, {Gavel}, {Goodsell}, {Graham},
  {Hartung}, {Hibon}, {Kalas}, {Konopacky}, {Larkin}, {Maire}, {Marchis},
  {McBride}, {Millar-Blanchaer}, {Morzinski}, {Norton}, {Oppenheimer},
  {Palmer}, {Patience}, {Perrin}, {Poyneer}, {Pueyo}, {Rantakyr{\"o}},
  {Sadakuni}, {Saddlemyer}, {Savransky}, {Soummer}, {Sivaramakrishnan}, {Song},
  {Thomas}, {Wallace}, {Wiktorowicz}, \& {Wolff}}]{Ingraham2014}
{Ingraham}, P., {Marley}, M.~S., {Saumon}, D., {et~al.} 2014, \apjl, 794, L15

\bibitem[{{Kalas} \& {Jewitt}(1995)}]{Kalas1995}
{Kalas}, P., \& {Jewitt}, D. 1995, Astronomical Journal, 110, 794

\bibitem[{{Kalas} {et~al.}(2008){Kalas}, {Graham}, {Chiang}, {Fitzgerald},
  {Clampin}, {Kite}, {Stapelfeldt}, {Marois}, \& {Krist}}]{Kalas2008}
{Kalas}, P., {Graham}, J.~R., {Chiang}, E., {et~al.} 2008, Science, 322, 1345

\bibitem[{{Kirkpatrick}(2005)}]{2005ARA&A..43..195K}
{Kirkpatrick}, J.~D. 2005, \araa, 43, 195

\bibitem[{{Kirkpatrick} {et~al.}(2006){Kirkpatrick}, {Barman}, {Burgasser},
  {McGovern}, {McLean}, {Tinney}, \& {Lowrance}}]{2006ApJ...639.1120K}
{Kirkpatrick}, J.~D., {Barman}, T.~S., {Burgasser}, A.~J., {et~al.} 2006, \apj,
  639, 1120

\bibitem[{Kirkpatrick {et~al.}(2008)Kirkpatrick, Cruz, Barman, Burgasser,
  Looper, Tinney, Gelino, Lowrance, Liebert, Carpenter, Hillenbrand, \&
  Stauffer}]{Kirkpatrick:2008ec}
Kirkpatrick, J.~D., Cruz, K.~L., Barman, T.~S., {et~al.} 2008, Astrophys. J.,
  689, 1295

\bibitem[{Kirkpatrick {et~al.}(2010)Kirkpatrick, Looper, Burgasser, Schurr,
  Cutri, Cushing, Cruz, Sweet, Knapp, Barman, Bochanski, Roellig, McLean,
  McGovern, \& Rice}]{Kirkpatrick:2010dc}
Kirkpatrick, J.~D., Looper, D.~L., Burgasser, A.~J., {et~al.} 2010, ApJS, 190,
  100

\bibitem[{{Lafreni{\`e}re} {et~al.}(2010){Lafreni{\`e}re}, {Jayawardhana}, \&
  {van Kerkwijk}}]{Lafreniere2010}
{Lafreni{\`e}re}, D., {Jayawardhana}, R., \& {van Kerkwijk}, M.~H. 2010,
  Astrophysical Journal, 719, 497

\bibitem[{{Lagrange} {et~al.}(2009){Lagrange}, {Gratadour}, {Chauvin}, {Fusco},
  {Ehrenreich}, {Mouillet}, {Rousset}, {Rouan}, {Allard}, {Gendron}, {Charton},
  {Mugnier}, {Rabou}, {Montri}, \& {Lacombe}}]{Lagrange2009}
{Lagrange}, A.-M., {Gratadour}, D., {Chauvin}, G., {et~al.} 2009, Astronomy and
  Astrophysics, 493, L21

\bibitem[{{Lagrange} {et~al.}(2010){Lagrange}, {Bonnefoy}, {Chauvin}, {Apai},
  {Ehrenreich}, {Boccaletti}, {Gratadour}, {Rouan}, {Mouillet}, {Lacour}, \&
  {Kasper}}]{Lagrange2010}
{Lagrange}, A.-M., {Bonnefoy}, M., {Chauvin}, G., {et~al.} 2010, Science, 329,
  57

\bibitem[{{Lagrange} {et~al.}(2012{\natexlab{a}}){Lagrange}, {Boccaletti},
  {Milli}, {Chauvin}, {Bonnefoy}, {Mouillet}, {Augereau}, {Girard}, {Lacour},
  \& {Apai}}]{2012A&A...542A..40L}
{Lagrange}, A.-M., {Boccaletti}, A., {Milli}, J., {et~al.} 2012{\natexlab{a}},
  \aap, 542, A40

\bibitem[{{Lagrange} {et~al.}(2012{\natexlab{b}}){Lagrange}, {Boccaletti},
  {Milli}, {Chauvin}, {Bonnefoy}, {Mouillet}, {Augereau}, {Girard}, {Lacour},
  \& {Apai}}]{Lagrange2012}
---. 2012{\natexlab{b}}, Astronomy and Astrophysics, 542, A40

\bibitem[{{Larkin} {et~al.}(2014){Larkin}, {Chilcote}, {Aliado}, {Bauman},
  {Brims}, {Canfield}, {Cardwell}, {Dillon}, {Doyon}, {Dunn}, {Fitzgerald},
  {Graham}, {Goodsell}, {Hartung}, {Hibon}, {Ingraham}, {Johnson}, {Kress},
  {Konopacky}, {Macintosh}, {Magnone}, {Maire}, {McLean}, {Palmer}, {Perrin},
  {Quiroz}, {Rantakyr{\"o}}, {Sadakuni}, {Saddlemyer}, {Serio}, {Thibault},
  {Thomas}, {Vallee}, \& {Weiss}}]{Larkin2014}
{Larkin}, J.~E., {Chilcote}, J.~K., {Aliado}, T., {et~al.} 2014, in \procspie,
  Vol. 9147, Ground-based and Airborne Instrumentation for Astronomy V, 91471K

\bibitem[{{Leggett} {et~al.}(2003){Leggett}, {Hawarden}, {Currie}, {Adamson},
  {Carroll}, {Kerr}, {Kuhn}, {Seigar}, {Varricatt}, \& {Wold}}]{Leggett2003}
{Leggett}, S.~K., {Hawarden}, T.~G., {Currie}, M.~J., {et~al.} 2003, \mnras,
  345, 144

\bibitem[{Liu {et~al.}(2013)Liu, Dupuy, \& Allers}]{Liu:2013ej}
Liu, M.~C., Dupuy, T.~J., \& Allers, K.~N. 2013, Astronomische Nachrichten,
  334, 85

\bibitem[{Liu {et~al.}(2016)Liu, Dupuy, \& Allers}]{Liu:2016co}
---. 2016, ApJ, 833, 96

\bibitem[{{Macintosh} {et~al.}(2006){Macintosh}, {Graham}, {Palmer}, {Doyon},
  {Gavel}, {Larkin}, {Oppenheimer}, {Saddlemyer}, {Wallace}, {Bauman}, {Evans},
  {Erikson}, {Morzinski}, {Phillion}, {Poyneer}, {Sivaramakrishnan}, {Soummer},
  {Thibault}, \& {Veran}}]{Macintosh2006}
{Macintosh}, B., {Graham}, J., {Palmer}, D., {et~al.} 2006, in Society of
  Photo-Optical Instrumentation Engineers (SPIE) Conference Series, Vol. 6272,
  Society of Photo-Optical Instrumentation Engineers (SPIE) Conference Series

\bibitem[{{Macintosh} {et~al.}(2014){Macintosh}, {Graham}, {Ingraham},
  {Konopacky}, {Marois}, {Perrin}, {Poyneer}, {Bauman}, {Barman}, {Burrows},
  {Cardwell}, {Chilcote}, {De Rosa}, {Dillon}, {Doyon}, {Dunn}, {Erikson},
  {Fitzgerald}, {Gavel}, {Goodsell}, {Hartung}, {Hibon}, {Kalas}, {Larkin},
  {Maire}, {Marchis}, {Marley}, {McBride}, {Millar-Blanchaer}, {Morzinski},
  {Norton}, {Oppenheimer}, {Palmer}, {Patience}, {Pueyo}, {Rantakyro},
  {Sadakuni}, {Saddlemyer}, {Savransky}, {Serio}, {Soummer},
  {Sivaramakrishnan}, {Song}, {Thomas}, {Wallace}, {Wiktorowicz}, \&
  {Wolff}}]{Macintosh2014}
{Macintosh}, B., {Graham}, J.~R., {Ingraham}, P., {et~al.} 2014, Proceedings of
  the National Academy of Science, 111, 12661

\bibitem[{{Macintosh} {et~al.}(2015){Macintosh}, {Graham}, {Barman}, {De Rosa},
  {Konopacky}, {Marley}, {Marois}, {Nielsen}, {Pueyo}, {Rajan}, {Rameau},
  {Saumon}, {Wang}, {Ammons}, {Arriaga}, {Artigau}, {Beckwith}, {Brewster},
  {Bruzzone}, {Bulger}, {Burningham}, {Burrows}, {Chen}, {Duchene}, {Esposito},
  {Fabrycky}, {Fitzgerald}, {Follette}, {Fortney}, {Gerard}, {Goodsell},
  {Greenbaum}, {Hibon}, {Hinkley}, {Hufford}, {Hung}, {Ingraham},
  {Johnson-Groh}, {Kalas}, {Lafreniere}, {Larkin}, {Lee}, {Line}, {Long},
  {Maire}, {Marchis}, {Matthews}, {Max}, {Metchev}, {Millar-Blanchaer},
  {Mittal}, {Morley}, {Morzinski}, {Murray-Clay}, {Oppenheimer}, {Palmer},
  {Patel}, {Patience}, {Perrin}, {Poyneer}, {Rafikov}, {Rantakyro}, {Rice},
  {Rojo}, {Rudy}, {Ruffio}, {Ruiz}, {Sadakuni}, {Saddlemyer}, {Salama},
  {Savransky}, {Schneider}, {Sivaramakrishnan}, {Song}, {Soummer}, {Thomas},
  {Vasisht}, {Wallace}, {Ward-Duong}, {Wiktorowicz}, {Wolff}, \&
  {Zuckerman}}]{Macintosh2015}
{Macintosh}, B., {Graham}, J.~R., {Barman}, T., {et~al.} 2015, Science, 350, 64

\bibitem[{{Maire} {et~al.}(2014){Maire}, {Ingraham}, {De Rosa}, {Perrin},
  {Rajan}, {Savransky}, {Wang}, {Ruffio}, {Wolff}, {Chilcote}, {Doyon},
  {Graham}, {Greenbaum}, {Konopacky}, {Larkin}, {Macintosh}, {Marois},
  {Millar-Blanchaer}, {Patience}, {Pueyo}, {Sivaramakrishnan}, {Thomas}, \&
  {Weiss}}]{2014SPIE.9147E..85M}
{Maire}, J., {Ingraham}, P.~J., {De Rosa}, R.~J., {et~al.} 2014, in \procspie,
  Vol. 9147, Ground-based and Airborne Instrumentation for Astronomy V, 914785

\bibitem[{{Males} {et~al.}(2014){Males}, {Close}, {Morzinski}, {Wahhaj}, {Liu},
  {Skemer}, {Kopon}, {Follette}, {Puglisi}, {Esposito}, {Riccardi}, {Pinna},
  {Xompero}, {Briguglio}, {Biller}, {Nielsen}, {Hinz}, {Rodigas}, {Hayward},
  {Chun}, {Ftaclas}, {Toomey}, \& {Wu}}]{Males2014}
{Males}, J.~R., {Close}, L.~M., {Morzinski}, K.~M., {et~al.} 2014,
  Astrophysical Journal, 786, 32

\bibitem[{{Marleau} {et~al.}(2017){Marleau}, {Klahr}, {Kuiper}, \&
  {Mordasini}}]{Marleau2017}
{Marleau}, G.-D., {Klahr}, H., {Kuiper}, R., \& {Mordasini}, C. 2017, ArXiv
  e-prints, arXiv:1701.02747

\bibitem[{Marley {et~al.}(2007)Marley, Fortney, Hubickyj, Bodenheimer, \&
  Lissauer}]{Marley:2007bf}
Marley, M.~S., Fortney, J.~J., Hubickyj, O., Bodenheimer, P., \& Lissauer,
  J.~J. 2007, ApJ, 655, 541

\bibitem[{{Marois} {et~al.}(2006){Marois}, {Lafreni{\`e}re}, {Doyon},
  {Macintosh}, \& {Nadeau}}]{Marois2006}
{Marois}, C., {Lafreni{\`e}re}, D., {Doyon}, R., {Macintosh}, B., \& {Nadeau},
  D. 2006, Astrophysical Journal, 641, 556

\bibitem[{{Marois} {et~al.}(2008){Marois}, {Macintosh}, {Barman}, {Zuckerman},
  {Song}, {Patience}, {Lafreni{\`e}re}, \& {Doyon}}]{Marois2008}
{Marois}, C., {Macintosh}, B., {Barman}, T., {et~al.} 2008, Science, 322, 1348

\bibitem[{{Marois} {et~al.}(2010){Marois}, {Zuckerman}, {Konopacky},
  {Macintosh}, \& {Barman}}]{Marois2010}
{Marois}, C., {Zuckerman}, B., {Konopacky}, Q.~M., {Macintosh}, B., \&
  {Barman}, T. 2010, Nature, 468, 1080

\bibitem[{{Mayor} \& {Queloz}(1995)}]{MayorQueloz1995}
{Mayor}, M., \& {Queloz}, D. 1995, Nature, 378, 355

\bibitem[{{Millar-Blanchaer} {et~al.}(2015){Millar-Blanchaer}, {Graham},
  {Pueyo}, {Kalas}, {Dawson}, {Wang}, {Perrin}, {moon}, {Macintosh}, {Ammons},
  {Barman}, {Cardwell}, {Chen}, {Chiang}, {Chilcote}, {Cotten}, {De Rosa},
  {Draper}, {Dunn}, {Duch{\^e}ne}, {Esposito}, {Fitzgerald}, {Follette},
  {Goodsell}, {Greenbaum}, {Hartung}, {Hibon}, {Hinkley}, {Ingraham},
  {Jensen-Clem}, {Konopacky}, {Larkin}, {Long}, {Maire}, {Marchis}, {Marley},
  {Marois}, {Morzinski}, {Nielsen}, {Palmer}, {Oppenheimer}, {Poyneer},
  {Rajan}, {Rantakyr{\"o}}, {Ruffio}, {Sadakuni}, {Saddlemyer}, {Schneider},
  {Sivaramakrishnan}, {Soummer}, {Thomas}, {Vasisht}, {Vega}, {Wallace},
  {Ward-Duong}, {Wiktorowicz}, \& {Wolff}}]{MillarBlanchaer2015}
{Millar-Blanchaer}, M.~A., {Graham}, J.~R., {Pueyo}, L., {et~al.} 2015, \apj,
  811, 18

\bibitem[{{Morzinski} {et~al.}(2015){Morzinski}, {Males}, {Skemer}, {Close},
  {Hinz}, {Rodigas}, {Puglisi}, {Esposito}, {Riccardi}, {Pinna}, {Xompero},
  {Briguglio}, {Bailey}, {Follette}, {Kopon}, {Weinberger}, \&
  {Wu}}]{Morzinski2015}
{Morzinski}, K.~M., {Males}, J.~R., {Skemer}, A.~J., {et~al.} 2015, \apj, 815,
  108

\bibitem[{{Mouillet} {et~al.}(1997){Mouillet}, {Larwood}, {Papaloizou}, \&
  {Lagrange}}]{Mouillet1997}
{Mouillet}, D., {Larwood}, J.~D., {Papaloizou}, J.~C.~B., \& {Lagrange}, A.~M.
  1997, Monthly Notices of the RAS, 292, 896

\bibitem[{{Nielsen} {et~al.}(2014){Nielsen}, {Liu}, {Wahhaj}, {Biller},
  {Hayward}, {Males}, {Close}, {Morzinski}, {Skemer}, {Kuchner}, {Rodigas},
  {Hinz}, {Chun}, {Ftaclas}, \& {Toomey}}]{Nielsen2014}
{Nielsen}, E.~L., {Liu}, M.~C., {Wahhaj}, Z., {et~al.} 2014, \apj, 794, 158

\bibitem[{{Oppenheimer} {et~al.}(2013){Oppenheimer}, {Baranec}, {Beichman},
  {Brenner}, {Burruss}, {Cady}, {Crepp}, {Dekany}, {Fergus}, {Hale},
  {Hillenbrand}, {Hinkley}, {Hogg}, {King}, {Ligon}, {Lockhart}, {Nilsson},
  {Parry}, {Pueyo}, {Rice}, {Roberts}, {Roberts}, {Shao}, {Sivaramakrishnan},
  {Soummer}, {Truong}, {Vasisht}, {Veicht}, {Vescelus}, {Wallace}, {Zhai}, \&
  {Zimmerman}}]{Oppenheimer2013}
{Oppenheimer}, B.~R., {Baranec}, C., {Beichman}, C., {et~al.} 2013,
  Astrophysical Journal, 768, 24

\bibitem[{{Perrin} {et~al.}(2014){Perrin}, {Maire}, {Ingraham}, {Savransky},
  {Millar-Blanchaer}, {Wolff}, {Ruffio}, {Wang}, {Draper}, {Sadakuni},
  {Marois}, {Rajan}, {Fitzgerald}, {Macintosh}, {Graham}, {Doyon}, {Larkin},
  {Chilcote}, {Goodsell}, {Palmer}, {Labrie}, {Beaulieu}, {De Rosa},
  {Greenbaum}, {Hartung}, {Hibon}, {Konopacky}, {Lafreniere}, {Lavigne},
  {Marchis}, {Patience}, {Pueyo}, {Rantakyr{\"o}}, {Soummer},
  {Sivaramakrishnan}, {Thomas}, {Ward-Duong}, \& {Wiktorowicz}}]{Perrin2014}
{Perrin}, M.~D., {Maire}, J., {Ingraham}, P., {et~al.} 2014, in \procspie, Vol.
  9147, Ground-based and Airborne Instrumentation for Astronomy V, 91473J

\bibitem[{{Pueyo}(2016)}]{Pueyo2016}
{Pueyo}, L. 2016, \apj, 824, 117

\bibitem[{{Quanz} {et~al.}(2010){Quanz}, {Meyer}, {Kenworthy}, {Girard},
  {Kasper}, {Lagrange}, {Apai}, {Boccaletti}, {Bonnefoy}, {Chauvin}, {Hinz}, \&
  {Lenzen}}]{Quanz2010}
{Quanz}, S.~P., {Meyer}, M.~R., {Kenworthy}, M.~A., {et~al.} 2010, \apjl, 722,
  L49

\bibitem[{{Rameau} {et~al.}(2013){Rameau}, {Chauvin}, {Lagrange}, {Boccaletti},
  {Quanz}, {Bonnefoy}, {Girard}, {Delorme}, {Desidera}, {Klahr}, {Mordasini},
  {Dumas}, \& {Bonavita}}]{Rameau2013}
{Rameau}, J., {Chauvin}, G., {Lagrange}, A.-M., {et~al.} 2013, Astrophysical
  Journal, Letters, 772, L15

\bibitem[{{Robert} {et~al.}(2016){Robert}, {Gagn{\'e}}, {Artigau},
  {Lafreni{\`e}re}, {Nadeau}, {Doyon}, {Malo}, {Albert}, {Simard}, {Bardalez
  Gagliuffi}, \& {Burgasser}}]{2016arXiv160706117R}
{Robert}, J., {Gagn{\'e}}, J., {Artigau}, {\'E}., {et~al.} 2016, ArXiv
  e-prints, arXiv:1607.06117

\bibitem[{{Schneider} {et~al.}(2014){Schneider}, {Cushing}, {Kirkpatrick},
  {Mace}, {Gelino}, {Faherty}, {Fajardo-Acosta}, \&
  {Sheppard}}]{2014AJ....147...34S}
{Schneider}, A.~C., {Cushing}, M.~C., {Kirkpatrick}, J.~D., {et~al.} 2014, \aj,
  147, 34

\bibitem[{Schneider {et~al.}(2016)Schneider, Windsor, Cushing, Kirkpatrick, \&
  Wright}]{Schneider:2016iq}
Schneider, A.~C., Windsor, J., Cushing, M.~C., Kirkpatrick, J.~D., \& Wright,
  E.~L. 2016, Astrophys. J., 822, L1

\bibitem[{{Smith} \& {Terrile}(1984)}]{SmithTerrile1984}
{Smith}, B.~A., \& {Terrile}, R.~J. 1984, Science, 226, 1421

\bibitem[{{Soummer} {et~al.}(2012){Soummer}, {Pueyo}, \&
  {Larkin}}]{Soummer2012}
{Soummer}, R., {Pueyo}, L., \& {Larkin}, J. 2012, Astrophysical Journal,
  Letters, 755, L28

\bibitem[{{Sparks} \& {Ford}(2002)}]{Sparks2002}
{Sparks}, W.~B., \& {Ford}, H.~C. 2002, Astrophysical Journal, 578, 543

\bibitem[{{Spiegel} \& {Burrows}(2012)}]{SpiegelBurrows2012}
{Spiegel}, D.~S., \& {Burrows}, A. 2012, \apj, 745, 174

\bibitem[{{Tokunaga} {et~al.}(2002){Tokunaga}, {Simons}, \&
  {Vacca}}]{2002PASP..114..180T}
{Tokunaga}, A.~T., {Simons}, D.~A., \& {Vacca}, W.~D. 2002, \pasp, 114, 180

\bibitem[{{van Leeuwen}(2007)}]{vanLeeuwen2007}
{van Leeuwen}, F. 2007, Astronomy and Astrophysics, 474, 653

\bibitem[{{Wagner} {et~al.}(2016){Wagner}, {Apai}, {Kasper}, {Kratter},
  {McClure}, {Robberto}, \& {Beuzit}}]{2016Sci...353..673W}
{Wagner}, K., {Apai}, D., {Kasper}, M., {et~al.} 2016, Science, 353, 673

\bibitem[{{Wang} {et~al.}(2014){Wang}, {Rajan}, {Graham}, {Savransky},
  {Ingraham}, {Ward-Duong}, {Patience}, {De Rosa}, {Bulger},
  {Sivaramakrishnan}, {Perrin}, {Thomas}, {Sadakuni}, {Greenbaum}, {Pueyo},
  {Marois}, {Oppenheimer}, {Kalas}, {Cardwell}, {Goodsell}, {Hibon}, \&
  {Rantakyr{\"o}}}]{wang2014}
{Wang}, J.~J., {Rajan}, A., {Graham}, J.~R., {et~al.} 2014, in \procspie, Vol.
  9147, Ground-based and Airborne Instrumentation for Astronomy V, 914755

\bibitem[{{Wang} {et~al.}(2016){Wang}, {Graham}, {Pueyo}, {Kalas},
  {Millar-Blanchaer}, {Ruffio}, {De Rosa}, {Ammons}, {Arriaga}, {Bailey},
  {Barman}, {Bulger}, {Burrows}, {Cardwell}, {Chen}, {Chilcote}, {Cotten},
  {Fitzgerald}, {Follette}, {Doyon}, {Duch{\^e}ne}, {Greenbaum}, {Hibon},
  {Hung}, {Ingraham}, {Konopacky}, {Larkin}, {Macintosh}, {Maire}, {Marchis},
  {Marley}, {Marois}, {Metchev}, {Nielsen}, {Oppenheimer}, {Palmer}, {Patel},
  {Patience}, {Perrin}, {Poyneer}, {Rajan}, {Rameau}, {Rantakyr{\"o}},
  {Savransky}, {Sivaramakrishnan}, {Song}, {Soummer}, {Thomas}, {Vasisht},
  {Vega}, {Wallace}, {Ward-Duong}, {Wiktorowicz}, \& {Wolff}}]{wang2016}
{Wang}, J.~J., {Graham}, J.~R., {Pueyo}, L., {et~al.} 2016, ArXiv e-prints,
  arXiv:1607.05272

\bibitem[{{Woitke} \& {Helling}(2003)}]{Woitke2003}
{Woitke}, P., \& {Helling}, C. 2003, \aap, 399, 297

\end{thebibliography}
\end{document}